\documentclass[aps,prb,twocolumn,superscriptaddress,floatfix,citeautoscript,hyperref=pdftex]{revtex4-1}

\newcommand{\be}{\begin{equation}}
\newcommand{\ee}{\end{equation}}

\usepackage[colorlinks=true]{hyperref}
\usepackage{bm}

\usepackage{times}
\usepackage{graphicx}
\usepackage{dcolumn}
\usepackage{bm}
\usepackage{color}
\usepackage{amsmath}
\usepackage{amssymb}
\usepackage{braket}

\begin{document}

\title{Incipient loop current order in the under-doped cuprate superconductors}
\author{Saheli Sarkar}
\author{Debmalya Chakraborty}
\author{Catherine P\'{e}pin}

\affiliation{Institut de Physique Th\'eorique, Universit\'e Paris-Saclay, CEA, CNRS, F-91191 Gif-sur-Yvette, France.}


\begin{abstract}
 There are growing experimental evidence which indicate discrete symmetry breaking like time-reversal ($\mathcal{T}$), parity ($\mathcal{P}$) and C$_{4}$ lattice rotation in the pseudo-gap state of the under-doped copper-oxide based (cuprate) superconductors. The discrete symmetry breaking manifests a true phase transition to an ordered state. A detailed thermodynamic understanding of these orders can answer various puzzles related to the nature of the transition at the pseudo-gap temperature T$^*$. In this work, we investigate thermodynamic signature of $\mathcal{T-P}$ symmetry breaking considering superconductivity (SC) and bond-density wave (BDW) as two primary orders. The BDW can generate both modulating charge and current densities. This framework takes into account an intricate competition between the ubiquitous charge density wave and SC, which is prominent in various cuprates in the under-doped regime. We demonstrate that within mean-field approach of competing BDW and SC orders, a $\mathcal{T-P}$ breaking ground state of coexisting BDW and SC can be stabilized, provided the BDW itself breaks $\mathcal{T}-\mathcal{P}$. But this ground state ceases to occur at higher temperatures. However, we show that  fluctuations in SC and BDW can drive emergence of a new unusual translational symmetry preserving order due to a preemptive phase transition by spontaneously breaking $\mathcal{T-P}$ at a higher temperature before the primary orders set in. We refer this order to be magneto-electric loop current (MELC) order. We present possible nature of phase transition for this new incipient MELC order and discuss some experimental relevance.
\end{abstract}


\maketitle
\section{Introduction}\label{intro}
The various anomalies in the normal state properties of under-doped regime of hole-doped cuprate superconductors have been a long standing puzzle in condensed matter physics. This peculiar normal state has a partially gapped Fermi surface and is known as `pseudo-gap' (PG) state. The PG state \cite{Alloul89,Warren89,Walstedt90,Alloul91,Walstedt91,berthier96,Marshall96,Harris97,Renner98,Ino98,tallon01,Ronning03,McElroy06} for the under-doped cuprates is usually set below a characteristic temperature T$^{*}$ well above the superconducting critical temperature T$_{c}$ as indicated in several experiments. Whether the PG state appears through a true phase-transition at T$^{*}$, associated with a symmetry breaking is still a debate. A complete understanding of the nature of this transition and the corresponding broken symmetries can unravel the mystery of the PG state.

\begin{figure}[t]
\includegraphics[width=\linewidth]{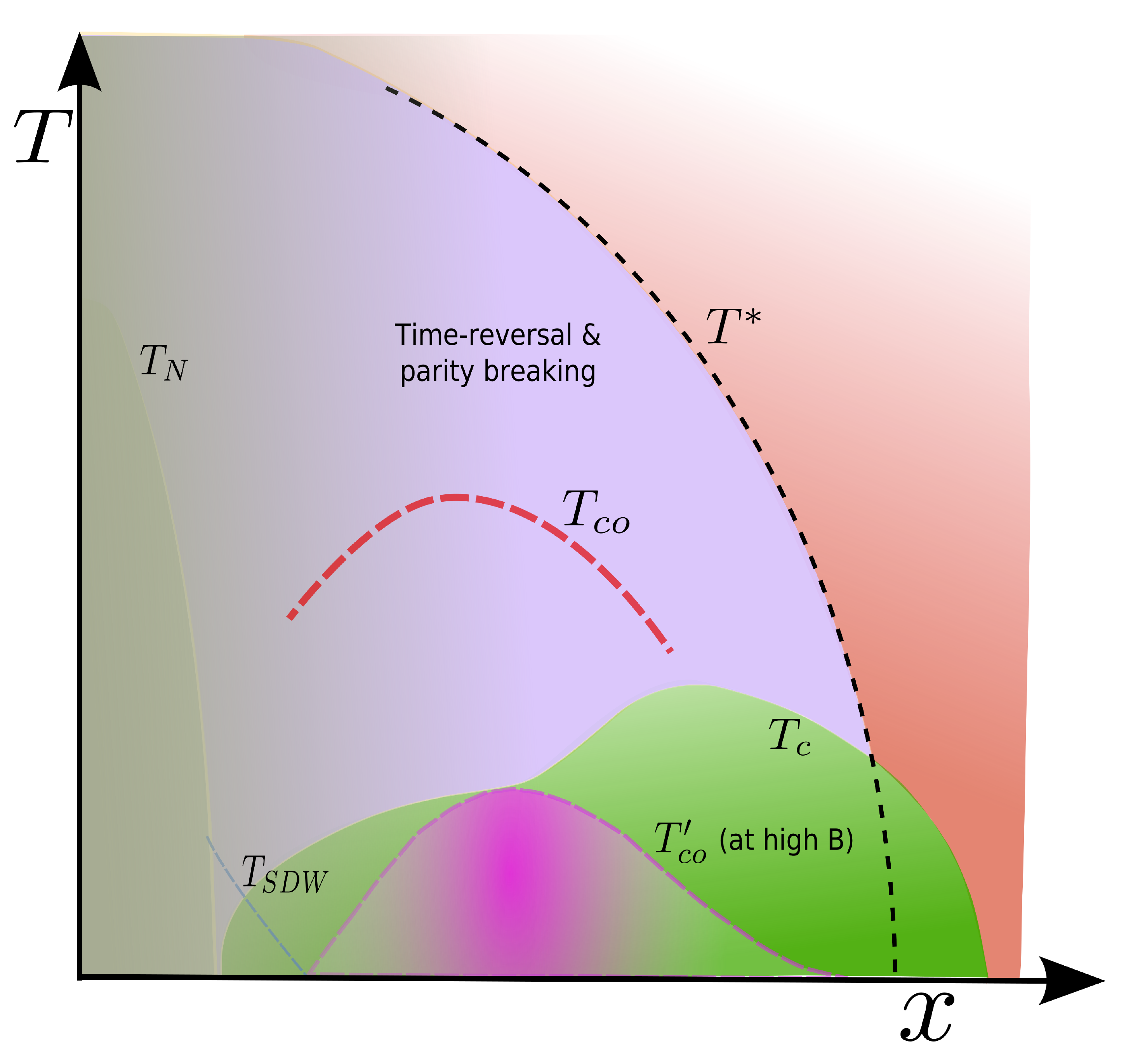}
\caption{\label{Fig_PhaseDiagrm}
A schematic temperature (T)- hole doping ($x$) phase diagram of cuprates summarizing experiments. Upon small doping, the system exhibits a d-wave superconducting phase below the critical temperature T$_{c}$ that follows a dome-like shape. A mysterious PG state appears below a temperature T$^{*}$, which is much higher than the T$_{c}$ in the under-doped region. Different experiments observed time-reversal and parity symmetry breaking in the PG state. A short-range CO is seen below T$_{co}$, whereas CO becomes long-range upon application of a magnetic field below temperature T$^{\prime}_{co}$.
}
\end{figure}

Several experiments have provided evidence of a true phase transition at T$^{*}$ and various symmetry breaking in the PG state. Ultrasound spectroscopy \cite{Shekhter13} and magnetic quantum oscillation measurements \cite{Ramshaw15} showed signatures of a thermodynamic phase transition at T$^{*}$. Moreover, recent experiments suggest breaking of discrete ($\mathbb{Z}_{2}$) symmteries in the PG state, which makes it intriguing to associate the discrete symmetry breaking  with the phase transition at T$^{*}$. Angle resolved photoemission spectroscopy (ARPES) with circularly polarized photons for under-doped Bi$_{2}$Sr$_{2}$CaCu$_{2}$O$_{8+\delta}$ (BSCCO)\cite{kaminski02} in the PG state suggested time-reversal symmetry breaking [Fig.\ref{Fig_PhaseDiagrm}]. Spin-polarized neutron scattering in YBa$_{2}$Cu$_{3}$O$_{6+x}$ (YBCO) \cite{Fauque06,Mook08}, HgBa$_{2}$CuO$_{4+\delta}$ \cite{li08,Li11} and Bi$_{2}$Sr$_{2}$CaCuO$_{8+\delta}$ \cite{DeAlmeida-Didry12} have shown evidence of long-range magnetic order at T$^{*}$ with wave-vector $\vec{Q}=0$. This  magnetic order preserves lattice translational invariance but breaks the time-reversal symmetry. Another spin polarized neutron scattering\cite{Baledent_10} in La$_{2-x}$Sr$_{x}$CuO$_{4}$ has reported short-range magnetic order. Optical second harmonic generation (OSHG) measurement\cite{Zhao2016} suggested breaking of parity symmetry at T$^{*}$ [Fig.\ref{Fig_PhaseDiagrm}]. Apart from these, polar Kerr effect measurements showed finite rotation of linearly polarized light reflected from the sample within the PG state in a number of under-doped  cuprates \cite{He11,Kapitulnik09,Karapetyan12}. The Kerr effect observations were interpreted in terms of time-reversal symmetry breaking\cite{Yakovenko:2015fd} and sometimes in terms of chiral symmetry breaking\cite{Karapetyan14,Aji13,Pershoguba13}. Scanning tunneling microscopy (STM) \cite{Lawler10,Kohsaka07}, anomalous Nernst effect \cite{Daou10}, torque magnetometry \cite{Sato:2017hg} and polarized neutron diffraction measurement\cite{Mangin_Thro17} have also detected nematic order inside the PG state which breaks the lattice C$_{4}$ rotational symmetry down to C$_{2}$, but preserves lattice translational symmetry  and hence is a $\vec{Q}=0$ order .

Signatures of time-reversal ($\mathcal{T}$) symmetry breaking from polarized neutron scattering experiments in different cuprate compounds \cite{Fauque06,li08,DeAlmeida-Didry12,Baledent_10} motivated several theoretical studies. Varma \textit{et.al} \cite{Varma97,Varma99,Simon02,Varma06} first proposed  the orgin of $\mathcal{T}$ breaking $\vec{Q}=0$ magnetic moments in the polarized neutron scattering in the PG state due to existence of Intra-Unit-Cell (IUC) loop currents. The IUC currents preserve lattice translational symmetry as they exist inside a single unit cell. The loop currents additionally break parity ($\mathcal{P}$) symmetry explaining the features of OSHG experiment \cite{Simon03}. The IUC loop currents proposed by Varma requires a three-band model of CuO$_2$ planes. The possibility of such currents in three-band microscopic models were further explored in several theoretical works \cite{Chudzinski_08,Fischer:2011du,Bulut:2015jt,Carvalho16,Atkinson_16,Scheurer18}. However, numerical analysis \cite{Greiter07,Thomale08} and quantum variational Monte-Carlo study\cite{Weber09} have challenged the existence and stability of such currents.

Furthermore, a nonsuperconducting $\vec{Q}=0$ order is not expected to open a gap on the Fermi surface, as it can either change the dispersion of the quasiparticle spectrum or can simply shift the chemical potential. Additionally it can not explain the presence of various $\vec{Q} \neq 0$ orders at low temperatures. These indicate that finite $\vec{Q}$ orders become imperative to open a gap on the Fermi surface, and indeed such orders have been observed in experiments. X-ray diffraction \cite{Ghiringhelli12,Chang12,wu12,Achkar12,Blackburn13a,Blackburn13b,Blanco-Canosa13,Croft14,daSilvaNeto:2014vy}and STM \cite{Hoffman02,matsuba07,Fujita14,machida16} ubiquitiously detected charge-density wave order (CO) for T$<$T$_{\text{CO}}$ [Fig.\ref{Fig_PhaseDiagrm}]. The CO breaks lattice translational symmetry and consequently has finite wave-vector ($\vec{Q} \neq 0$). There are also experimental indications \cite{Hamidian15a,Comin15,Achkar12,Achkar13a,Kohsaka07} that the modulations of the charge density lives primarily on Cu-O-Cu bonds. In presence of high magnetic field, the CO attains a long range nature in YBCO\cite{Wu11,Wu13a,Chang:2016gz,Gerber:2015gx,Jang16,leboeuf13,Laliberte2018} below T$^{\prime}_{\text{CO}}$ [Fig.\ref{Fig_PhaseDiagrm}]. It is also interesting to note that, various experiments \cite{Blackburn13a,Ghiringhelli12,Wu:2015bt,Hoffman02,Hamidian16} observed competition between CO and SC. More recently, local Josephson scanning tunneling microscopy have provided evidence of a distinct finite $\vec{Q}$ electronic order, namely a pair-density-wave (PDW)\cite{larkin65,Fulde:1964dq,Agterberg19} in BSCCO \cite{Hamidian16,Edkins18}.

Although there are no signatures of long-range finite $\vec{Q}$ orders close to T$^{*}$, experiments hint towards presence of various fluctuations. This motivated several theoretical proposals on fluctuations of different order parameters driving the phenomenology of the PG state. Proximity to Mott localization transition \cite{Lee06} inspired scenarios with superconducting phase fluctuations \cite{EmeryVJ:1995dr,Banerjee:2011bz} and fluctuating preformed Cooper pairs \cite{Norman:1998va,Chien2009,Boyack18}. On the other hand, presence of various competing orders \cite{Fradkin:2015ch,Keimer:2015vy} in the under-doped region of the phase diagram led to theories with fluctuations in various orders like spin density waves \cite{Chatterjee17,Sachdev19}, staggered flux phase \cite{Wen96,Lee:1998cr}, CO \cite{Efetov13,Metlitski10,Wang14} and PDW \cite{Wang18,Dai18,Dai19,Norman18}. There are also proposals based on fluctuations guided by an emergent symmetry \cite{Zhang97,Efetov13,Metlitski10}. The emergent SU(2) theory describes some of the phenomenology \cite{Meier13,Meier14,Einenkel14,Hayward14,Chowdhury:2014cp,Kloss15,Montiel16,Kloss16,Morice:2017kd,Chakraborty18} for the PG phase. Recent experimental signatures of fluctuations in both SC\cite{Rajasekaran18,Hu14,Fausti11} and charge\cite{Loret19} channels in the PG state motivated a theoretical proposal\cite{Chakraborty19} for pseudo-gap based on entangled fluctuating preformed particle-particle and particle-hole pairs. While particle-particle pairs result to the SC state, the particle-hole pairs give rise to bond-density wave (BDW) order which can result into both charge wave order and current density wave. 

Fluctuating orders can also be considered as possible candidates for explaining the antinodal gap on the Fermi surface in the PG phase. For instance, superconducting fluctuations can result into Fermi arcs as observed in ARPES for T$>$T$_{c}$ because the nodal quasi particles are more prone to thermal fluctuations than the anti-nodal ones \cite{Campuzano98}. Within the fluctuation scenarios, therefore it is of fundamental importance to investigate discrete symmetry breaking $\vec{Q}=0$ orders close to temperature T$^{*}$ and its connection to the PG transition. Some phenomenological works\cite{Wang14,Agterberg:2014wf} discussed discrete symmetry breaking in the PG state using composite CO and PDW fields. But no general consensus, whether these theoretical frameworks can consolidate the mechanism of pseudo-gap and the discrete symmetry breaking orders, has been reached so far.

In this paper, we are interested in the following question: whether fluctuations in both SC and BDW hold the key to the $\mathcal{T}-\mathcal{P}$ symmetry breaking in the PG state resulting in a phase transition at T$^{*}$. Towards this, we first investigate a Ginzburg-Landau (GL) theory of competing primary BDW and SC orders at the mean-field level without considering fluctuations. We notice that $\mathcal{T}-\mathcal{P}$ symmetry can only be broken in a coexisting ground state of SC and BDW, where BDW itself breaks $\mathcal{T}-\mathcal{P}$ symmetry. Such a ground state has never been observed in experiments. This immediately drives the possibility of a role played by the fluctuations in the SC and BDW. To analyze the effect of fluctuations, we construct a composite PDW order from higher order combinations of primary SC and BDW orders. The composite PDW order has the same wave-vector $\vec{Q}$ as that of BDW and has same charge as SC. The fluctuations in both SC and BDW lead to the fluctuations in composite PDW order. Using a Hubbard-Stratonovich (HS) method with saddle-point approximation in a GL theory of fluctuating composite PDW order, we find a nontrivial order which can break $\mathcal{T}-\mathcal{P}$ in the PG state. In our phenomenological treatment, the $\mathcal{T}-\mathcal{P}$ symmetry breaking order parameter shares same symmetry properties as that of the IUC magneto-electric loop current (MELC) proposed by Varma. Hence, we call this order parameter as MELC order in the rest of this paper. However, the MELC order in our theory is an emergent $\vec{Q}=0$ order, formed by a higher order combination of primary SC and BDW, and thus does not need resorting to the three-band models.

We organize the rest of the paper as follows. In section \ref{sec OP} we define the two primary order parameters, SC and BDW, and construct a composite PDW order parameter. We discuss in detail the symmetry properties of the order parameters in the context of cuprates. In section \ref{sec:Au_MELC} we build an auxiliary MELC order parameter and discuss its symmetry transformation properties. In section \ref{sec:GL_nofluc} we present the mean-field GL theory of competing SC and BDW and calculate the analytic conditions for obtaining $\mathcal{T}-\mathcal{P}$ breaking ground state. In section \ref{sec:GL_fluc} we present the HS approach for fluctuations and detail the possible nature of phase transition depending on parameters, through which the preemptive MELC ground state can appear. In section \ref{Sec_Disc} we put forward discussions on the extensions of the framework and possible relevance to experimental findings in cuprates. In section \ref{conclusion} we conclude by presenting a summary of our work.

\section{Order parameters: symmetry properties}\label{sec OP}
\begin{figure}[t]
\includegraphics[width=0.7\linewidth]{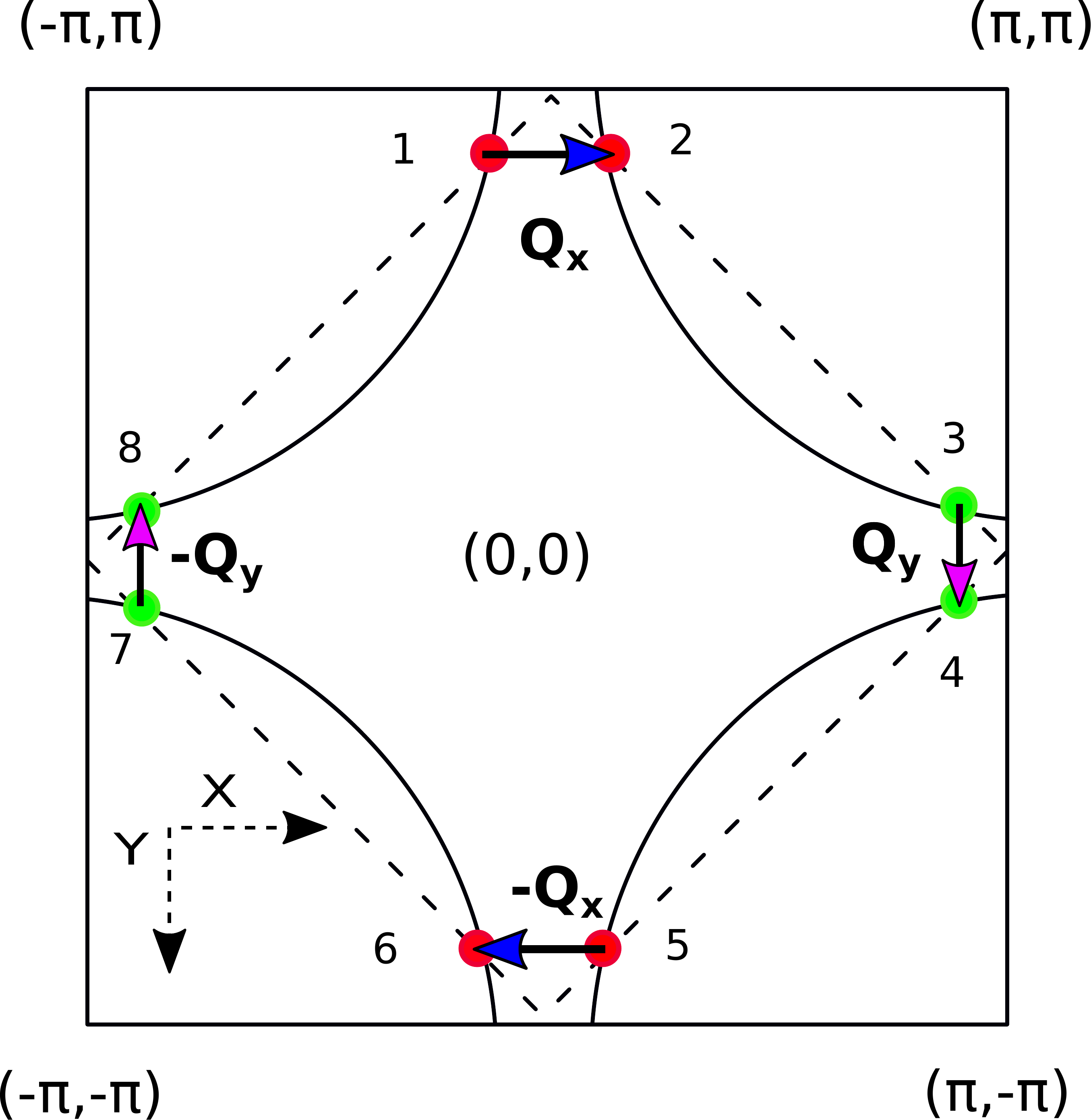}
\caption{\label{Fig_FS}A schematic representation of the Fermi surface of hole doped cuprates. The dotted lines represent the magnetic Brillouin zone boundary and the solid curved lines denote the Fermi surface. The Fermi surface intersects the magnetic Brillouin zone boundary at eight hot-spots, which are shown by red and green dots and numbered from 1 to 8. As mentioned in the text, we consider the BDW order parameters with axial wave vectors connecting the nearest hot spots in the first Brillouin zone. The BDW wave vectors for hot spots 1,3,5 and 7 are shown by the arrows. For orthorhombic lattice systems, BDW order parameters corresponding to only the red hot-spots are relevant. On the other hand, for tetragonal lattice systems, BDW order parameters corresponding to both red and green hots spots have to be considered for writing the free energy.
}
\end{figure}

In this section, we introduce the Fermi surface [Fig.\ref{Fig_FS}] and the order parameters of a typical hole-doped cuprate. More specifically, we discuss here two primary order parameters, BDW and SC and the secondary order parameter, composite PDW and their symmetry transformation properties. 
 \begin{table*}[t]
\begin{center}
 \begin{tabular}{||c | c c c ||c c c||} 
 \hline
\textbf{Point group} & & \textbf{Primary orders} & & ~~~~~~~~\textbf{Composite orders} & & \\
 \textbf{operations} & $\chi_{Q}^{k}$ & $\chi_{-Q}^{-k}$ & $\Delta$ &$\Phi_{Q}^{k}$ & $\Phi_{-Q}^{-k}$ & \\
 \hline\hline
$R_{x}$: &  $\chi_{Q}^{k}$ & $\chi_{-Q}^{-k}$ & $\Delta$ &$\Phi_{Q}^{k}$ & $\Phi_{-Q}^{-k}$ & \\
\hline
$R_{y}$: &  $\chi_{-Q}^{-k}$ & $\chi_{Q}^{k}$ & $\Delta$ &$\Phi_{-Q}^{-k}$ & $\Phi_{Q}^{k}$ & \\
\hline
$R_{z}$: &  $\chi_{-Q}^{-k}$ & $\chi_{Q}^{k}$ & $\Delta$ &$\Phi_{-Q}^{-k}$ & $\Phi_{Q}^{k}$ & \\
\hline
$\sigma_{x}$: &  $\chi_{-Q}^{-k}$ & $\chi_{Q}^{k}$ & $\Delta$ &$\Phi_{-Q}^{-k}$ & $\Phi_{Q}^{k}$ & \\
\hline
$\sigma_{y}$: &  $\chi_{Q}^{k}$ & $\chi_{-Q}^{-k}$ & $\Delta$ &$\Phi_{Q}^{k}$ & $\Phi_{-Q}^{-k}$ & \\
\hline
$\sigma_{z}$: &  $\chi_{Q}^{k}$ & $\chi_{-Q}^{-k}$ & $\Delta$ &$\Phi_{Q}^{k}$ & $\Phi_{-Q}^{-k}$ & \\
 \hline
\end{tabular}
\caption{Point group symmetry transformation of SC, BDW and PDW fields in orthorhombic lattice. Here $R_{x}, R_{y}$ and $R_{z}$ represent the action of the two-fold (C$_{2}$) rotations about x, y and z-axis respectively. $\sigma_{x}, \sigma_{y}$ and $\sigma_{z}$ represent action of mirror reflections about $y-z$, $z-x$ and $x-y$ planes respectively.\label{symmetry_op}}
\end{center}
\end{table*}

Underdoped cuprates \cite{Metlitski10b,Efetov13} are often described by a spin-fermion model \cite{Abanov00,abanov03,Abanov01} based on antiferromagnetic fluctuations. Within this scenario, the PG phase is described by an emergent symmetry between the SC order and a BDW order with diagonal wave-vectors connecting different `hot-spots' [labeled as 1-8 in Fig.\ref{Fig_FS}], where the Fermi surface intersects the magnetic Brillouin zone boundary \cite{Efetov13}. But experimentally the CO is observed with wave vectors either horizontal or vertical to the crystallographic axes \cite{daSilvaNeto14,Comin14}. The magnitude of the wave vector is found to be very close to the axial wave vector connecting two neighboring `hot-spots' \cite{Comin14}. Theoretically, BDW with axial wave vectors can also be obtained as one of the competing instabilities in models with antiferromagnetic fluctuations \cite{Wang14}. The BDW with axial wave-vector can be enhanced by including fluctuations\cite{Wang14,Meier14,Chowdhury:2014cp}, considering dynamic exchange interactions \cite{Wang15c} or an offsite Coulomb interaction\cite{Allais14c} in the microscopic models. Furthermore, a recent work\cite{Chakraborty19} including both antiferromagnetic interactions and off-site Coulomb repulsion shows that a BDW with axial wave-vector dominates near the hot-spots on the Fermi surface. Motivated by all these theoretical indications and experimental observations, we consider the BDW order parameter only at the hot-spots with axial wave vectors  ($\vec{Q}_{x}$ or $\vec{Q}_{y}$) as shown in the Fig.\ref{Fig_FS}.

The complex BDW order parameter $\chi_{Q}^{k}$ with ordering wave vector $\vec{Q}$ at each given momentum ($k$) is given as $\sum_{\sigma} \langle c^{\dagger}_{k+Q,\sigma}c_{k,\sigma}\rangle$. In the special case where $\chi_{Q}^{k}$ describes only current modulations with $Q=(\pi,\pi)$, the corresponding order parameter is often referred to as d-density waves (dDW) \cite{Chakravarty01} or staggered flux order parameter \cite{Affleck88,Marston89,Hsu91} in the literature. While the dDW ground state itself breaks $\mathcal{T}-\mathcal{P}$ \cite{Chakravarty01}, the BDW ground state considered in this paper does not necessarily break $\mathcal{T}-\mathcal{P}$, as we will see in sections \ref{sec:GL_nofluc} and \ref{sec:GL_fluc} (also see Ref. \onlinecite{Wang14}).

The complex superconducting order parameter $\Delta$ is given by $\langle c^{\dagger}_{k,\uparrow}c^{\dagger}_{-k,\downarrow}\rangle$. Next we introduce the PDW order parameter, which is composite of $\chi_{Q}^{k}$ and $\Delta$ and can be defined as,
\begin{align}\label{eq:Phi_def}
\Phi_{Q}^{k} = \chi_{Q}^{k}\Delta.
\end{align}
We would like to emphasize that the wave-vector of the PDW order $\Phi_{Q}^{k}$ is the same as that of BDW wave-vector. A different set of theories \cite{Lee14,Dai18,Wang18,agterberg08} considered fluctuating primary PDW, albeit with a different wave-vector $\vec{P}$ from that of the CO wave-vector.

For tetragonal crystal systems, all the hot-spot pairs (1-2, 3-4, 5-6, and 7-8) need to be considered because of C$_{4}$ symmetry in the system. In this case, there are eight complex BDW order parameters for each hot-spot point. For orthorhombic systems, the relevant hot-spot pairs are 1-2 and 5-6 as the $C_{4}$ symmetry is now absent and one can not bring 1-2 and 5-6 pairs to any of the hot-spot pairs 7-8 or 3-4. Therefore, in orthorhombic case, there are four complex order parameters corresponding to the hot-spot points 1, 2, 5 and 6 [shown with red dots in Fig.\ref{Fig_FS}]. The complex order parameters for orthorhombic system, corresponding to the primary order manifold are,
\begin{align}\label{eq:OM_1}
[\chi_{Q_{x}}^{1},\chi_{-Q_{x}}^{2},\chi_{-Q_{x}}^{5},\chi_{Q_{x}}^{6},\Delta].
\end{align}
To build the Ginzburg-Landau free energy density, the point-group symmetry transformation properties of these order parameters are required. For orthorhombic case, the required  point group symmetries are: three two-fold axes of rotations about x, y and z axis and three mirror planes: y-z, z-x and x-y. We are interested only in the parity and (or) time-reversal symmetry breaking. Thus, we do not consider any mirror symmetry breaking ground states. As a result, without any loss of generality, we implement the following two equalities,
\begin{align}\label{eq:sigmay}
\nonumber
\chi_{Q_{x}}^{1} &= \chi_{Q_{x}}^{6} \equiv  \chi_{Q}^{1} \\
\chi_{-Q_{x}}^{5} &= \chi_{-Q_{x}}^{2} \equiv \chi_{-Q}^{5}.
\end{align}
Subsequently the order parameter space is further reduced to a smaller subset,
\begin{align}\label{eq:sigmay_2}
[\chi_{Q}^{1},\chi_{-Q}^{5},\Delta].
\end{align}
Rewriting $\chi_{Q}^{1}$ and $\chi_{-Q}^{5}$ as $\chi_{Q}^{k}$ and $\chi_{-Q}^{-k}$ respectively, the order parameter manifold becomes,
\begin{align}\label{eq:sigmay_3}
[\chi_{Q}^{k}, \chi_{-Q}^{-k},\Delta].
\end{align}

Finally, we summarize the point group symmetry transformation of the BDW, SC and composite PDW in the table \ref{symmetry_op}. The order parameter manifold for the tetragonal lattice systems will be larger than in Eq.~\eqref{eq:sigmay_2} and the number of possible ground states will be increased. While we explicitly show the order parameter manifold and discuss the case for tetragonal systems in section \ref{sec_om}, we will restrict the analysis in the rest of this paper for orthorhombic lattice systems with no mirror symmetry breaking order parameters, in order to have an analytical control in a reduced parameter space.

Under parity and time-reversal the BDW and the composite PDW transform as follows,
\begin{align}\label{eq:sym_or1}
\begin{split}
\chi_{Q}^{k} \xrightarrow{\mathcal{P}} \chi_{-Q}^{-k}, \\
\chi_{Q}^{k} \xrightarrow{\mathcal{T}} \chi_{-Q}^{\dagger -k},\\
\Phi_{Q}^{k} \xrightarrow{\mathcal{P}} \Phi_{-Q}^{-k},\\
\Phi_{Q}^{k} \xrightarrow{\mathcal{T}} \Phi_{-Q}^{\dagger -k}.
\end{split}
\end{align}
For convenience from now on we will suppress the $k$ index from the BDW order parameters, for example we will use $\chi_{Q}$ and $\chi_{-Q}$ for $\chi_{Q}^{k}$ and $\chi_{-Q}^{-k}$ respectively. The same notation will be also applied to the composite PDW.
\section{Loop current order}\label{sec:Au_MELC}
Our goal in this work is to study the $\mathcal{T-P}$ symmetry breaking due to emergence of an order in the PG state. Here, we define such an order parameter, referred as MELC order, which is translationally invariant ($\vec{Q} = 0$).  Within our theoretical framework, this order appears to be an auxiliary order. The concept of auxiliary orders have been introduced previously in several contexts \cite{Fernandes12,Wang14,Agterberg:2014wf,Fradkin:2015ch,Scheurer18}, and sometimes they are referred  to as `vestigial'  or `secondary' orders. 

In the similar spirit, we construct the MELC order parameter, $\ell$ from the primary BDW and SC order parameters and is given by the following equation, 
\begin{align}\label{eq:LC_def1}
\ell = |\chi_{Q}\Delta|^{2} - |\chi_{-Q}\Delta|^{2}.
\end{align}
Equivalently, $\ell$ can be written in terms of the composite PDW order parameters using Eq.\eqref{eq:Phi_def} as follows
\begin{align}\label{eq:LC_def2}
\ell = |\Phi_{Q}|^{2} - |\Phi_{-Q}|^{2}.
\end{align}
Upon time-reversal and parity transformation, the BDW and composite PDW transform as given by the Eq.\eqref{eq:sym_or1}. Using Eq.\eqref{eq:sym_or1}, $\ell$ transforms under time-reversal and parity as,
\begin{align}\label{eq:lc1}
\ell \xrightarrow{\mathcal{T}} -\ell, \ell  \xrightarrow{\mathcal{P}} -\ell, \ell  \xrightarrow{\mathcal{T}\mathcal{P}} \ell.
\end{align}
This depicts that the order parameter $\ell$ breaks the time-reversal, parity but preserves their product. Under a spatial translation by {$\vec{R}$}, $\chi_{Q}$ and $\Phi_{Q}$ transform as $\chi_{Q} \rightarrow e^{i\vec{Q}.\vec{R}}\chi_{Q}$ and $\Phi_{Q} \rightarrow e^{i\vec{Q}.\vec{R}}\Phi_{Q}$ respectively . Hence $\ell$ remains invariant under a spatial translation $\vec{R}$ and therefore is a $\vec{Q}=0$ order. Magneto-electric IUC loop currents proposed by Varma \cite{Varma97} also have similar transformation properties under time-reversal, parity and spatial translation.
\begin{table*}[t]
\begin{center} 
 \begin{tabular}{||c c c | c| c | c c |c||} 
 \hline
& \textbf{Primary orders} & & \textbf{Broken symmetry} & \textbf{Free energy} & ~~~~~~~~\textbf{Composite PDW} & & \textbf{MELC} \\ 
$\chi_{Q}$ & $\chi_{-Q}$ & $\Delta$ & \textbf{in ground state} & &$\Phi_{Q}$ & $\Phi_{-Q}$ & $\ell$ \\ 
 \hline\hline
a & 0 & b & $ U(1) \times U(1) \times \mathbb{Z}_{2}$ &$F (a,0,b) = \frac{\left[2\alpha\alpha_{d}\beta_{2}- \alpha_{d}^{2}\beta_{1}- \alpha^{2}\beta_{d}\right]}{2\left[\beta_{1}\beta_{d} - \beta_{2}^{2}\right]}$ & ab & 0  & $a^{2}b^{2}$  \\ 
 \hline
0 & a & b & $ U(1) \times U(1) \times \mathbb{Z}_{2}$ & $F (0,a,b) = \frac{\left[2\alpha\alpha_{d}\beta_{2}- \alpha_{d}^{2}\beta_{1}- \alpha^{2}\beta_{d}\right]}{2\left[\beta_{1}\beta_{d} - \beta_{2}^{2}\right]}$& 0 & ab  & -$a^{2}b^{2}$ \\
 \hline
  a & a & b &$ U(1) \times U(1)$  & $F(a,a,b) = \frac{4\alpha\alpha_{d}\beta_{2} - \alpha_{d}^{2}(\beta_{1}+\beta_{3})-2\alpha^{2}\beta_{d}}{2\beta_{d}(\beta_{1}+\beta_{3})-4\beta_{2}^{2}}$& ab & ab & 0 \\
 \hline
  a & 0 & 0 &$ U(1) \times \mathbb{Z}_{2}$ & $F (a,0,0) = \frac{-\alpha^{2}}{2\beta_{1}}$ & 0 & 0  & 0 \\
 \hline
  0 & a & 0 &$ U(1) \times \mathbb{Z}_{2}$ & $F (0,a,0) = \frac{-\alpha^{2}}{2\beta_{1}}$ & 0 & 0  & 0 \\
 \hline
  0 & 0 & b &$ U(1)$ & $F(0,0,b) = \frac{-\alpha_{d}^{2}}{2\beta_{d}}$ & 0 & 0 & 0 \\
 \hline
  a & a & 0 &$ U(1)$ & $F(a, a, 0) = \frac{-\alpha^{2}}{(\beta_{1} + \beta_{3})}$& 0 &0  & 0 \\  
 \hline
\end{tabular}
\caption{Table showing the properties of all possible mean-field ground states. The non-zero values of $\chi_{Q}$ and $\chi_{-Q}$ in these first two columns are same as the free energy density Eq.\eqref{eq_GLFE} is invariant under $\chi_{Q} \leftrightharpoons\chi_{-Q}$. The fourth column shows the symmetries broken by the corresponding ground state of the primary orders. The fifth column gives the free energy for each state. The last three columns show the composite PDW order and the MELC order constructed from the primary orders. Only the states (a, 0, b) and (0, a, b) can sustain a non-zero value of the MELC order.\label{tstaes}}
\end{center}
\end{table*}

\section{Mean-field Ginzburg-Landau theory of SC and BDW}\label{sec:GL_nofluc}
This section aims to investigate the formation of the MELC order by constructing a mean-field Ginzburg-Landau theory of competing BDW and SC orders. The GL free energy density functional for a spatially homogeneous case, which remains invariant under translations, time-reversal, parity, gauge symmetries as well as all the point group symmetry operations of the orthorhombic system, is given as follows,
\begin{widetext}
\begin{align}\label{eq_GLFE}
\nonumber
F &= \alpha_{d}|\Delta|^{2} + \alpha\left(|\chi_{Q}|^{2} + |\chi_{-Q}|^{2}\right) + \frac{\beta_{1}}{2}\left(|\chi_{Q}|^{4} + |\chi_{-Q}|^{4}\right) 
+ \frac{\beta_{d}}{2}|\Delta|^{4} \\
&+ \beta_{2}\left(|\chi_{Q}|^{2}|\Delta|^{2} + |\chi_{-Q}|^{2}|\Delta|^{2}\right) 
+ \beta_{3}\left(|\chi_{Q}|^{2}|\chi_{-Q}|^{2}\right) 
  +\beta_{4}\left[\chi_{Q}\chi_{-Q}|\Delta|^{2} + |\Delta|^{2}(\chi_{Q}\chi_{-Q})^{*}\right].
\end{align}  
\end{widetext}
In the free energy Eq.\eqref{eq_GLFE}, $\beta_{2}$ gives the coupling between BDW field $\chi_{Q}$ or $\chi_{-Q}$ and superconducting field $\Delta$. $\beta_{3}$ represents coupling between $\chi_{Q}$ and $\chi_{-Q}$. And lastly $\beta_{4}$ gives the mutual coupling between the three fields $\chi_{Q}$, $\chi_{-Q}$ and $\Delta$. In this work, we are particularly interested in a ground state which spontaneously breaks only $\mathcal{T-P}$ symmetry and can sustain a MELC. Subsequently we replace $\beta_{4} = 0$ in the free energy density for simplification which excludes possibilities of pure imaginary ground states. 
\subsection{All possible ground states }\label{MF_sub1}
We now discuss the possible mean-field ground states of SC and BDW orders from the free energy Eq.\eqref{eq_GLFE}. The calculations to obtain the solutions for the ground states by minimizing the free energy  Eq.\eqref{eq_GLFE} and the resulting free energy of the respective ground states are shown in the appendix \ref{sec:AppA}. Table \ref{tstaes} summarizes the seven possible mean-field ground states of the free energy Eq.\eqref{eq_GLFE} and the free energy for each ground state. For $\alpha <0$, and $\beta_{1}>0$, the state (a, 0, 0) becomes a minimum, where the field $\chi_{Q}$ condenses with $\chi_{Q} \neq 0$. The superconducting state (0, 0, b) becomes a minimum with $\Delta\neq 0$ when $\alpha_{d} <0$, and $\beta_{d}>0$. 
We note that, there can be another BDW state (a, a, 0) present, when both $\chi_{Q}$ and $\chi_{-Q}$ are non-zero and superconducting order parameter is absent. These two fields condense when $\alpha <0$ and $(\beta_{1}+\beta_{3})>0 $.

We note in table \ref{tstaes}, that the two states (a, 0, b) and (0, a, b), where $|\chi_{Q}| \neq |\chi_{-Q}|$, have a $\mathbb{Z}_{2}$ degeneracy as their free energies are equal. This degeneracy can be lifted by spontaneously breaking the $\mathcal{T-P}$ symmetry and such a state will sustain a finite MELC, as can be seen from the last column of the table. No other states can sustain a finite MELC, which can also be seen from the table. Henceforth, we analyze the conditions on the GL parameters for the state (a, 0, b) to be the most stable ground state.

\subsection{Stability conditions for $(a,0,b)$ ground state}
Now we analyze the stability for the state (a, 0, b). The ground state (a,0,b) has a free energy, 
\begin{equation} \label{ew:fr_fc1_MainTxt}
F (a,0,b) = \frac{\left[2\alpha\alpha_{d}\beta_{2}- \alpha_{d}^{2}\beta_{1}- \alpha^{2}\beta_{d}\right]}{2\left[\beta_{1}\beta_{d} - \beta_{2}^{2}\right]}.
\end{equation}
The stability conditions are essentially the conditions for which the state (a, 0, b) is a global minimum. These conditions are given by,
\begin{subequations}
\begin{align}
 F(a,0,b) - F(a,a,b) < 0, \label{eq:stability_a}\\
 F(a,0,b) - F(a,0,0) < 0,\label{eq:stability_b}\\
 F(a,0,b) - F(0,0,b) < 0,\label{eq:stability_c}\\
 F(a,0,b) - F(a,a,0) < 0 \label{eq:stability_d}.
\end{align} 
\end{subequations}
\begin{figure*}[t]
\includegraphics[width=0.8\textwidth]{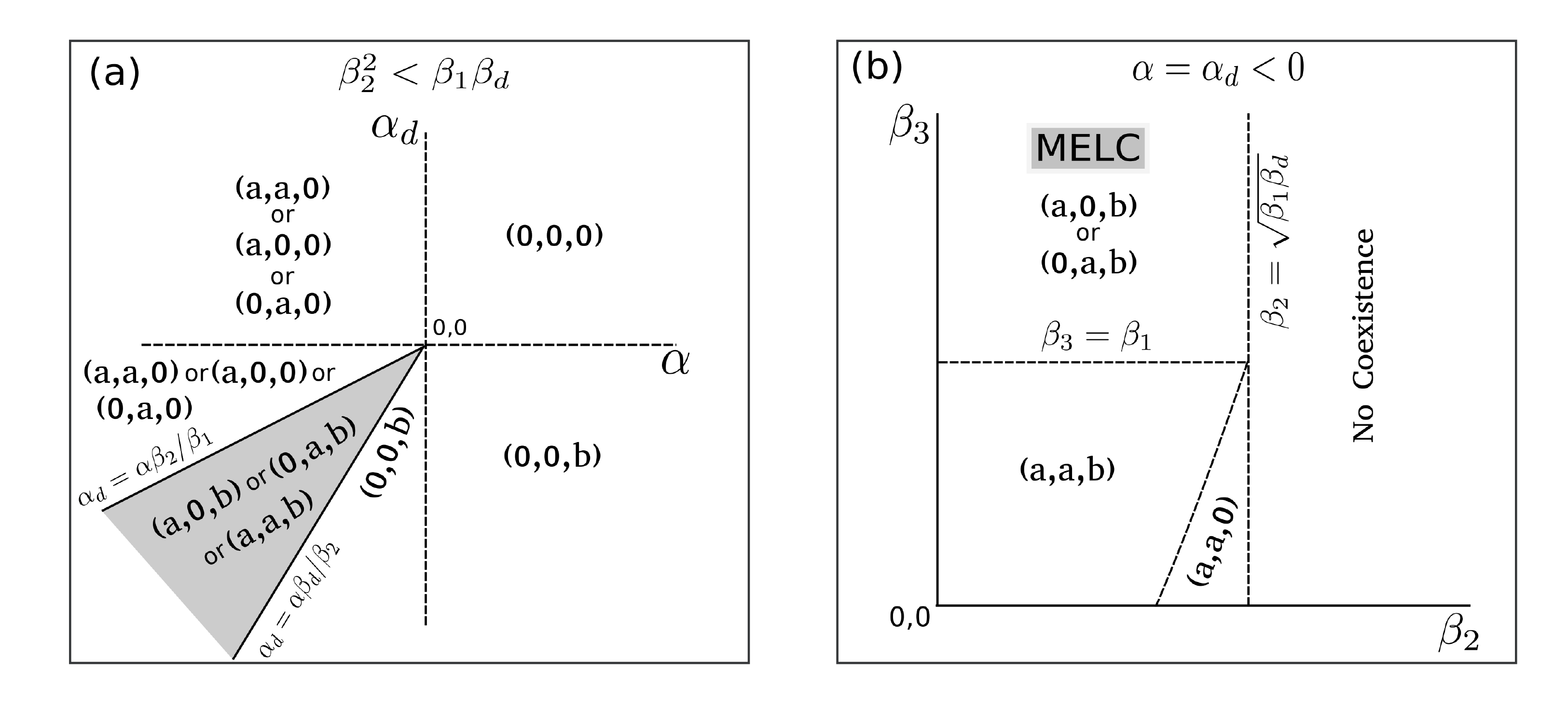}
\caption{\label{Fig_Phasediag_MF} Phase diagrams showing the parameter regimes for the occurrence of a MELC ground state obtained from the mean-field solutions of the GL free energy Eq.\eqref{eq_GLFE}. \textbf{(a)} Illustration of different phases in the parameter space of $\alpha$ and $\alpha_{d}$: We choose $\beta_{2}^{2} < \beta_{1}\beta_{d}$ as it is a necessary condition for a coexistence of SC and BDW orders. For any positive values of both $\alpha$ and $\alpha_{d}$, the system is in a disordered state (0,0,0). If $\alpha_{d} <0$ and $\alpha >0$, only possible ground state is SC state (0,0,b). If $\alpha_{d} > 0$ and $\alpha < 0$, the possible ground states are (a,a,0), (a,0,0) and (0,a,0), where SC order vanishes and only BDW orders exist. When both of $\alpha_{d} < 0$ and $\alpha < 0$, coexistent states of SC and BDW orders emerge as indicated by the grey colored region. This region of coexistence is bounded by two lines obtained from Eq.\eqref{eq:mass1} and Eq.\eqref{eq:mass2}. There are three possible coexistent states: (a,0,b), (0,a,b) and (a,a,b). However, only (a,0,b) or (0,a,b) support a MELC order. \textbf{(b)} Role of the GL parameter $\beta_{3}$: With a choice of $\alpha=\alpha_{d}<0$ [inside the grey region in (a)], the MELC ground state is found to be bounded below by $\beta_{3} = \beta_{1}$. It is also seen that when $\beta_{3} < \beta_{1}$, the state (a,a,b) wins over MELC ground state, as lower value of $\beta_{3}$ favors existence two BDW orders $\chi_{Q}$ and $\chi_{-Q}$. Also the existence of state (a,a,0) is found from comparing free energies of state (a,a,b) and (a,a,0). In the parameter regime of $\beta_{2} > \sqrt{\beta_{1}\beta_{d}}$ no coexistence of BDW and SC orders can be found.}
\end{figure*}
Thus the stable ground state is achieved by the simultaneous fulfillment of the above conditions provided all the other minima of the GL free energy exist. The minimum requirements for the existence of all the minima are $\alpha<0$, $\alpha_{d}<0$, $\beta_{1}>0$, $\beta_{1}+\beta_{3}>0$ and $\beta_{d}>0$, which we derived in \ref{MF_sub1}. These conditions for the parameters will be valid for the rest of the discussion in this section. 

A close inspection of the comparison of free energies [Eqs.\eqref{eq:stability_a},\eqref{eq:stability_b},\eqref{eq:stability_c} and \eqref{eq:stability_d}] will also give insight to the strength of relative coupling between various fields.
Towards this, we first evaluate the condition for which Eq.\eqref{eq:stability_b} holds. This imposes  the following additional constraint on the coupling constant $\beta_{2}$ as 
\begin{align}\label{eq:coexst}
\left(\beta_{2}^{2} - \beta_{1}\beta_{d}\right) <0
\end{align}
and imposes two more conditions on the masses:
\begin{align}\label{eq:mass1}
\beta_{2}\alpha_{d} > \alpha\beta_{d},
\end{align}
 and 
\begin{align} \label{eq:mass2}
 \alpha\beta_{2} > \alpha_{d}\beta_{1}.
 \end{align}
From Eq.\eqref{eq:stability_c}, we get the same criteria as of Eqs.\eqref{eq:mass1} and \eqref{eq:mass2}. The condition Eq.\eqref{eq:stability_d} gives,
\begin{align}\label{eq:cond(a,a,0)}
2\left(\beta_{1} + \beta_{3}\right)\left(2\alpha\alpha_{d}\beta_{2} - \alpha_{d}^{2}\beta_{1} -\alpha^{2}\beta_{d}\right) > 2\alpha^{2}\left(\beta_{1}\beta_{d} -\beta_{2}^{2}\right).
\end{align}
Finally we investigate the stability criteria Eq.\eqref{eq:stability_a}. The free energy density for the state (a, a, b) is given by,
\begin{equation}\label{eq:frEn_(a,a,b)}
F(a,a,b) = \frac{4\alpha\alpha_{d}\beta_{2} - \alpha_{d}^{2}(\beta_{1}+\beta_{3})-2\alpha^{2}\beta_{d}}{2\beta_{d}(\beta_{1}+\beta_{3})-4\beta_{2}^{2}}.
\end{equation}
 We notice from the free energy density Eq.\eqref{eq:frEn_(a,a,b)}, that there are two conditions for the state (a, a, b) to become one of the possible minima , i.e.  $F(a,a,b) < 0$. The two conditions are given by the following equations,
\begin{subequations}\label{eq:cond(a,a,b)}
\begin{align}
2\beta_{d}\left(\beta_{1} + \beta_{3}\right) - 4\beta_{2}^{2} < 0, \\
4\alpha\alpha_{d}\beta_{2} - \alpha_{d}^{2}(\beta_{1} + \beta_{3}) - 2\alpha^{2}\beta_{d} > 0,
\end{align}
\end{subequations}
or, 
\begin{subequations}
\begin{align}
2\beta_{d}\left(\beta_{1} + \beta_{3}\right) - 4\beta_{2}^{2} > 0, \\
4\alpha\alpha_{d}\beta_{2} - \alpha_{d}^{2}(\beta_{1} + \beta_{3}) - 2\alpha^{2}\beta_{d} < 0.
\end{align} 
\end{subequations}
For the state (a, 0, b) to become more stable over the state (a, a, b) the condition Eq.\eqref{eq:stability_a} has to be satisfied. This poses an additional constraint
\begin{align}\label{eq:cr:final_free1}
\begin{split}
2(\beta_{1}\beta_{d} - \beta_{2}^{2})\left[4\alpha\alpha_{d}\beta_{2} - \alpha_{d}^{2}(\beta_{1} + \beta_{3}) - 2\alpha^{2}\beta_{d}\right] \\
 < \left[2(\beta_{1}+\beta_{3})\beta_{d} - 4\beta_{2}^{2}\right]\left(2\alpha\alpha_{d}\beta_{2} - \alpha_{d}^{2}\beta_{1} -\alpha^{2}\beta_{d}\right)
 \end{split}
\end{align}
on the masses and coupling constants in the free energy Eq.\eqref{eq_GLFE}.

In Fig.\ref{Fig_Phasediag_MF}, we present phase diagrams showing all the possible ground states and highlight the parameter regime where a ground state sustaining the MELC order is stable. As already indicated earlier in this section, only a state with coexisting SC and BDW fields can give rise to the MELC order. From Eq.\eqref{eq:coexst}, such a state is allowed only when $\beta_{2}^{2}<\beta_{1}\beta_{d}$. Restricting $\beta_{2}$ in this regime, in Fig.\ref{Fig_Phasediag_MF}(a), we plot the phase diagram in the parameter space of $\alpha$ and $\alpha_{d}$. The three possible coexistent states (a, 0, b), (0, a, b) and (a, a, b) are stable only in the grey region as shown in Fig.\ref{Fig_Phasediag_MF}(a). But as we saw in the table \ref{tstaes}, $\ell$ can have a non-zero value only for (a, 0 ,b) and (0, a, b). This imposes further condition on other GL parameters. To further illustrate, we consider a particular line $\alpha=\alpha_{d}<0$ in the grey region of Fig.\ref{Fig_Phasediag_MF}(a) and investigate the GL parameter $\beta_{3}$ in Fig.\ref{Fig_Phasediag_MF}(b). We find that (a, 0, b) and (0, a, b) are stable when $\beta_{3}>\beta_{1}$. 

However, the two states (a, 0, b) and (0, a, b) have a $\mathbb{Z}_{2}$ degeneracy due to the presence of $\mathcal{T-P}$ symmetry. Subsequently, the ground states of composite PDW also have the same $\mathbb{Z}_{2}$ degeneracy. This degeneracy can be lifted if the BDW itself spontaneously breaks $\mathcal{T-P}$ symmetry. As a consequence, the $\mathcal{T-P}$ broken ground state will have a finite MELC order as can be seen from Eq.\eqref{eq:LC_def1}.
\section{Fluctuating orders and preemptive MELC order}\label{sec:GL_fluc}
The $\mathcal{T-P}$ breaking mean-field BDW ground state, which we discussed in the section \ref{sec:GL_nofluc} has not yet been observed and such a ground state can not persist at higher temperatures. This raises the possibility of role played by fluctuations in $\mathcal{T-P}$ breaking. In this section, we therefore analyze fluctuation effects of BDW and SC on $\mathcal{T}-\mathcal{P}$ breaking in the PG state.

The fluctuations in both BDW and SC introduce gradient terms of both SC and BDW orders in the GL free energy Eq.\eqref{eq_GLFE}. This gives rise to additional number of parameters in the free energy Eq.\eqref{eq_GLFE}. As a result, it becomes more complex to study the free energy analytically. But, we recall from section \ref{sec:Au_MELC}, that MELC order parameter is defined as $\ell = |\chi_{Q}\Delta|^{2} - |\chi_{-Q}\Delta|^{2}$ , where the $\chi_{Q}\Delta$ and $\chi_{-Q}\Delta$ are nothing but two composite PDW fields $\Phi_{Q}$ and $\Phi_{-Q}$ respectively. Hence to investigate the $\mathcal{T}-\mathcal{P}$ breaking, alternatively we can write the free energy in terms of fluctuating composite PDW order parameters. This allows us to perform a systematic analytical study of the $\mathcal{T}-\mathcal{P}$ symmetry breaking in the PG state. The free energy in terms of fluctuating PDW orders will have the same symmetry properties as that of free energy in terms of SC and BDW in section \ref{sec:GL_nofluc}, as PDW and BDW transform in the exact same fashion under all the symmetry transformation as discussed in section \ref{symmetry_op}.

Before considering the fluctuations, we write the GL free energy for the composite PDW in a homogeneous system and incorporating all the symmetries for a disordered normal state of the system. The corresponding free energy is given by,
{\small{
\begin{align}\label{eq:MF_GL_phi_1}
\begin{split}
F_{0}[\Phi_{Q},\Phi_{-Q}] &= \alpha_{\phi} \left(|\Phi_{Q}|^{2} + |\Phi_{-Q}|^2\right) + \beta_{\varrho}\left(|\Phi_{Q}|^{4} + |\Phi_{-Q}|^{4}\right) \\
&+ 2\beta |\Phi_{Q}|^{2}|\Phi_{-Q}|^{2},
\end{split}
\end{align}}}
where $\beta$ and $\beta_{\varrho}$ are positive.
Next rescaling the parameters and rearranging the terms in the free energy Eq.\eqref{eq:MF_GL_phi_1}, we get
\begin{align}\label{eq:MF_GL_phi_2}
\begin{split}
F_{0}[\Phi_{Q},\Phi_{-Q}] &= \alpha^{\prime} \left(|\Phi_{Q}|^{2} + |\Phi_{-Q}|^2\right) + \frac{1}{2}\left(|\Phi_{Q}|^{2} + |\Phi_{-Q}|^{2}\right)^{2} \\
&- \frac{\beta_{\ell}}{2} \left(|\Phi_{Q}|^{2}|-|\Phi_{-Q}|^{2}\right)^{2},
\end{split}
\end{align}
where $\alpha^{\prime} = \alpha_{\phi}/(\beta_{\varrho}+\beta)$ and $\beta_{\ell} = (\beta-\beta_{\varrho})/(\beta_{\varrho}+\beta)$. $\beta_{\varrho}$ is the self interaction of both the fields $\Phi_{Q}$ and $\Phi_{-Q}$ and $\beta$ denotes the strength of the competition between the two fields. The rescaled parameter $\beta_{\ell}$ decides the nature of the PDW ground state. If $\beta<\beta_{\varrho}$ [i.e. $\beta_{\ell}<0$], the free energy in Eq.\eqref{eq:MF_GL_phi_1} favors a coexisting ground state with $(\Phi_{Q} \ne 0, \Phi_{-Q} \ne 0)$. On the other hand, $\beta>\beta_{\varrho}$ [i.e. $\beta_{\ell}>0$] allows for a ground state with $(\Phi_{Q} \ne 0, \Phi_{-Q}= 0)$ or $(\Phi_{-Q} \ne 0, \Phi_{Q}= 0)$. These two states have a $\mathbb{Z}_{2}$ degeneracy due to $\mathcal{T-P}$ symmetry, which is the same $\mathbb{Z}_{2}$ symmetry in the primary BDW order. 

Now, we add the gradient terms in $F_{0}$ [Eq.\eqref{eq:MF_GL_phi_2}] to account for the thermal fluctuations of the composite PDW field in a 2D system and arrive at the following GL free energy:
{\small{
\begin{align}\label{F2_CPDW}
\begin{split}
F[\Phi_{Q},\Phi_{-Q}] &= \alpha^{\prime} \left(|\Phi_{Q}|^{2} + |\Phi_{-Q}|^2\right) + \frac{1}{2}\left(|\Phi
_{Q}|^{2} + |\Phi_{-Q}|^{2}\right)^{2} \\
&-\frac{\beta_{\ell}}{2}\left(|\Phi_{Q}|^{2} - |\Phi_{-Q}|^{2}\right)^{2} + \left(|\nabla \Phi_{Q}|^{2} + |\nabla \Phi_{-Q}|^{2}\right).
\end{split}
\end{align}}}
Incorporating the fluctuations in $\Phi_{Q}$ and $\Phi_{-Q}$, through Hubbard-Stratonovich transformations of the GL free energy Eq.\eqref{F2_CPDW} and using saddle-point approximation, we study the possibility of appearance of $\mathcal{T-P}$ symmetry breaking without  any requirement of $\mathcal{T}-\mathcal{P}$ symmetry breaking in the PDW ground state.

\subsection{HS transformations and effective free energy}\label{HS_sub1}
The partition function corresponding to the free energy Eq.\eqref{F2_CPDW} can be written as $Z[\Phi_{Q},\Phi_{-Q}] \propto \int d\Phi_{Q} d\Phi_{-Q} \exp(-F[\Phi_{Q}, \Phi_{-Q}])$. We make a HS transformation of the partition function $Z[\Phi_{Q},\Phi_{-Q}]$, to arrive at an effective partition function in terms of the HS fields. To this end, we introduce two conjugate HS fields as follows, 
\begin{align}\label{eq:HSfields}
\begin{split}
\varrho & \equiv i (|\Phi_{Q}|^{2} + |\Phi_{-Q}|^{2}),\\
\ell & \equiv (|\Phi_{Q}|^{2} - |\Phi_{-Q}|^{2}).
 \end{split}
\end{align} 
The HS field $\ell$ describes a `preemptive' MELC order. $\varrho$ gives the Gaussian correction to susceptibility due to fluctuation. Now we make the following HS transformations in the partition function $Z[\Phi_{Q},\Phi_{-Q}]$:
\begin{widetext}
\begin{align}
\nonumber
\exp\left[-\sum_{i=1}^{N}(|\Phi_{Q,i}|^{2} + |\Phi_{-Q,i}|^{2})^{2}/2N \right]& = \sqrt{N/2\pi}\int d\varrho e^{\frac{-N\varrho^{2}}{2}}\exp\left[i\varrho \sum_{i=1}^{N}(|\Phi_{Q,i}|^{2} + |\Phi_{-Q,i}|^{2})\right] ,\\
\nonumber
\exp\left[\sum_{i=1}^{N}\beta_{\ell}(|\Phi_{Q,i}|^{2} - |\Phi_{-Q,i}|^{2})^{2}/2N \right]& = \sqrt{N/2\pi}\int d\ell e^{\frac{-N\ell^{2}}{2\beta_{\ell}}}\exp\left[\ell \sum_{i=1}^{N}(|\Phi_{Q,i}|^{2} - |\Phi_{-Q,i}|^{2})\right],
\end{align}
\end{widetext}
where we have assumed that the PDW fields have $N$ components, where $N\gg1$. Next we take the limit $N \sim 1$, where the qualitative results for HS transformation is not expected to change \cite{Tsvelik14,Wang14}. Using the above HS transformations with $N=1$ in the partition function $Z[\Phi_{Q},\Phi_{-Q}]$, and integrating over $\Phi_{Q}$ and $\Phi_{-Q}$, we obtain an effective partition function in terms of the HS fields $\varrho$ and $\ell$. The new effective partition can be written in terms of HS fields as
\begin{align*}
\begin{split}
Z_{eff}[\varrho,\ell] & \propto \int d\varrho d\ell \exp\left[-\frac{\varrho^{2}}{2} - \frac{\ell^{2}}{2\beta_{\ell}}\right]\\& \exp\left[-\int \frac{d^{2}q}{4\pi^{2}} \ln \left[(\alpha^{\prime} + q^{2} - i\varrho)^{2} -\ell^{2}\right]\right].
\end{split}
\end{align*}
As the effective partition function can be written as $Z_{eff}[\varrho,\ell]\propto \int d\varrho d\ell \exp(-F_{eff}[\varrho,\ell])$, the effective free energy $F_{eff}[\varrho,\ell]$ is given by,
\begin{align}\label{eqn_free_HS_eff}
F_{eff}[\varrho,\ell] = \frac{\varrho^{2}}{2} + \frac{\ell^{2}}{2\beta_{\ell}} + \int \frac{d^{2}q}{4\pi^{2}} \ln \left[(\alpha^{\prime} + q^{2} - i\varrho)^{2} -\ell^{2}\right].
\end{align}

\subsection{Saddle-point analysis of the effective free energy}\label{Saddle_sub1}

We consider that the fluctuations in the preemptive MELC order $\ell$ around the saddle-point solutions are small. Therefore we continue with the saddle-point approximation for the free energy in terms of the MELC order parameter and closely follow the theoretical framework in Refs.~[\onlinecite{Fernandes12}] and [\onlinecite{Wang14}].
 
The saddle point solutions are obtained by minimizing the free energy $F_{eff}[\varrho,\ell]$ w.r.t. $\varrho$ and $\ell$. These give the following equations for $\varrho$ and $\ell$,
\begin{align}\label{eq:part_rho}
\begin{split}
\frac{\partial F_{eff}}{\partial \varrho} &= 0 \\
\Rightarrow\varrho &= 2 i \int \frac{d^{2}q}{4\pi^{2}}\frac{(\alpha^{\prime} + q^{2} - i\varrho)}{(\alpha^{\prime} + q^{2} - i\varrho)^{2} -\ell^{2}}.
\end{split}
\end{align}
and,
\begin{align}\label{eq:part_lc}
\begin{split}
\frac{\partial F_{eff}}{\partial \ell} &= 0 \\
\Rightarrow\ell &= 2 \beta_{\ell} \int \frac{d^{2}q}{4\pi^{2}}\frac{\ell}{(\alpha^{\prime} + q^{2} - i\varrho)^{2} -\ell^{2}}.
\end{split}
\end{align}
After performing the integration in Eqs.\eqref{eq:part_rho} and \eqref{eq:part_lc}, we get the following two coupled equations for $r$ and $\ell$, 
\begin{align}\label{eq:rho_l}
\begin{split}
r &= \alpha^{\prime} + \frac{1}{4\pi}\left[\ln(\Lambda^{2}-\ell^{2})- \ln(r^{2} - \ell^{2})\right],\\ 
\ell &= \frac{\beta_{\ell}}{2\pi} \coth^{-1}(\frac{r}{\ell}),
\end{split}
\end{align}
where $r = \alpha^{\prime} - i\varrho$ and $\Lambda$ is the upper momentum cut-off. We note that the solution of Eq. \eqref{eq:part_rho} exists only for imaginary $\varrho$, hence we replace $\varrho$ by $i \varrho_{0}$ where $\varrho$ is real.
So, $\varrho = 0$ can not be a solution. But $\ell =0$ is an allowed solution. Therefore, we consider the following two cases: $\ell = 0, \varrho \neq 0$ and $\ell \neq 0, \varrho \neq 0$.
\subsubsection{Case: $\ell = 0, \varrho \neq 0$}
For the case $\ell = 0, \varrho \neq 0$, the solution for $r$ can be rewritten from Eq.\eqref{eq:rho_l} as,
\begin{align*}
r = \alpha^{\prime} + \frac{1}{2\pi}\ln\left(\frac{\Lambda}{r}\right).
\end{align*}
To find the stability of the solution, we need to analyze the condition for $\frac{\partial^{2} F_{eff}[\varrho,\ell]}{\partial \varrho^{2}}|_{\varrho= i \varrho_{0},\ell =0}$ and $\frac{\partial^{2} F_{eff}[\varrho,\ell]}{\partial \ell^{2}}|_{\varrho= i \varrho_{0},\ell =0}$ to be positive. We get the second derivatives to be as follows,
\begin{align}\label{eq:2ndderF_1}
\frac{\partial^{2} F_{eff}[\varrho,\ell]}{\partial \varrho^{2}}|_{\ell =0} = 1 + 2 i \int \frac{d^{2}q}{4\pi^{2}}\frac{-i}{(\alpha^{\prime} + q^{2} - i\varrho)^{2}}
\end{align}
and
\begin{align}\label{eq:2ndderF_2}
\frac{\partial^{2} F_{eff}[\varrho,\ell]}{\partial \ell^{2}} = \frac{1}{\beta_{\ell}}
 - 2  \int \frac{d^{2}q}{4\pi^{2}}\left[\frac{1}{(\alpha^{\prime} + q^{2} - i\varrho)^{2}-\ell^{2}} \right.
\nonumber \\\left. -\frac{2\ell^{2}}{[(\alpha^{\prime} + q^{2} - i\varrho)^{2}-\ell^{2}]^{2}}\right].
\end{align}
\begin{figure}[t]
\includegraphics[width=0.8\linewidth]{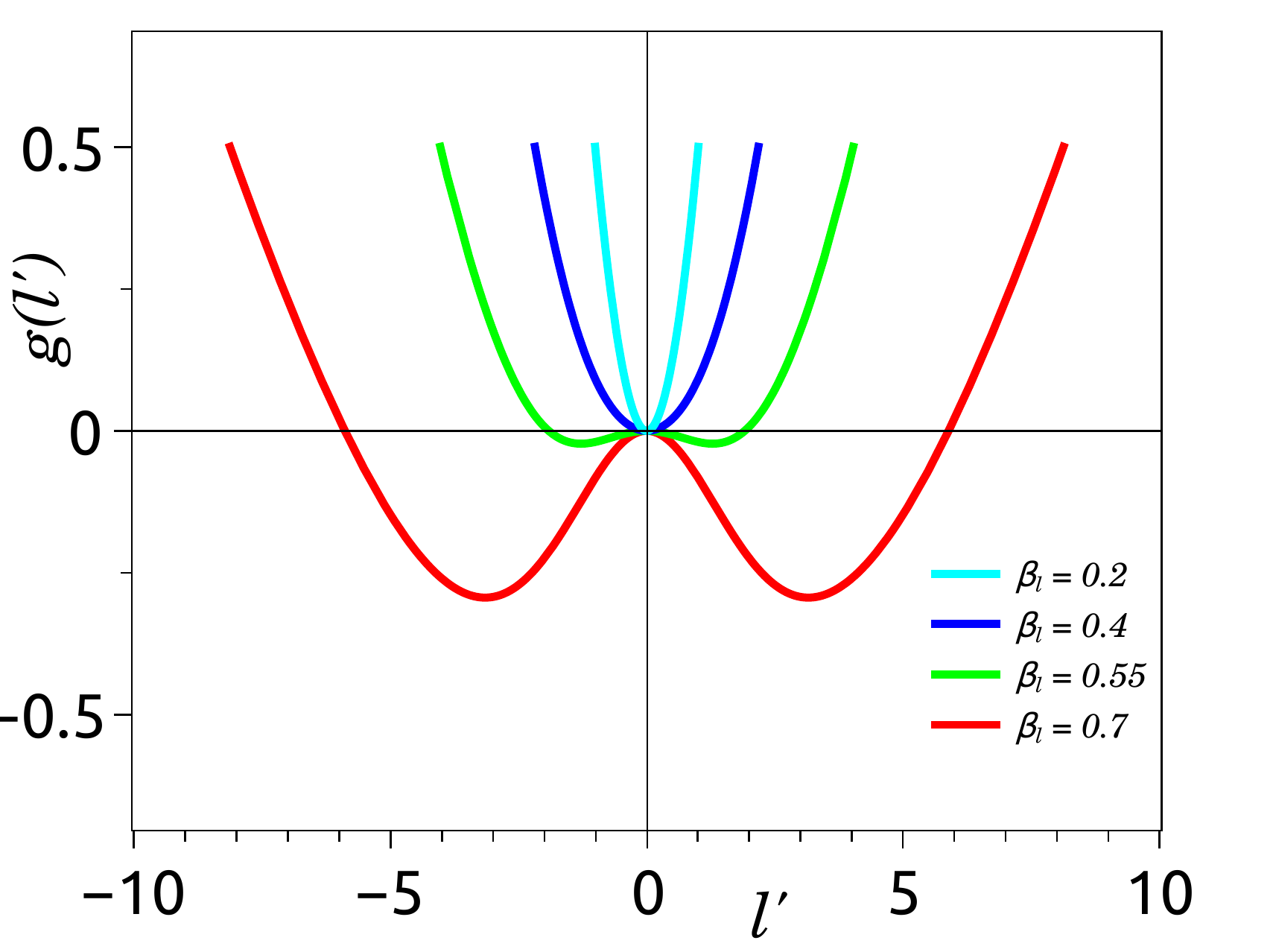}
\caption{\label{Plot_compare_betas}Plots of $g(\ell^{\prime})$ [defined in the Eq.\eqref{eq:def_g}] representing the L.H.S. of  Eq.\eqref{eq:func_l} which gives the solution for the rescaled MELC order $\ell^{\prime}$ [defined in Eq.\eqref{eq:l_rescale}] for four different values of $\beta_{\ell}$. In the range $0<\beta_{\ell}<0.5$, $g(\ell^{\prime})$ shows only one minimum at $\ell^{\prime}=0$. On the other hand, in the range $0<\beta_{\ell}<0.5$, $g(\ell^{\prime})$ acquires two minima for $\ell^{\prime} \ne 0$, symmetric about $\ell^{\prime}=0$.}
\end{figure}
Next plugging in $\varrho = i \varrho_{0}$ in Eq.\eqref{eq:2ndderF_1} and performing the integration
gives $\frac{\partial^{2} F_{eff}[\varrho,\ell]}{\partial \varrho^{2}}|_{\varrho= i \varrho_{0},\ell =0}=\left(1 + \frac{1}{2\pi r}\right)$. Here, we used $r = \alpha^{\prime} + \varrho_{0}$. The sign of $r$ is always positive unless the PDW fields become ordered. Hence the value of $\left(1 + \frac{1}{2\pi r}\right)$ is always positive.
Again, performing the integration and plugging in the limits $\varrho= i \varrho_{0}$ and $\ell =0$ in Eq.\eqref{eq:2ndderF_2}, we get,$
\frac{\partial^{2} F_{eff}[\varrho,\ell]}{\partial \ell^{2}}\vert_{\varrho= i \varrho_{0},\ell =0} = \frac{1}{\beta_{\ell}}(1-\frac{\beta_{\ell}}{2\pi r})$.
Now to hold $\frac{\partial^{2} F_{eff}[\varrho,\ell]}{\partial \ell^{2}}|_{\varrho= i \varrho_{0},\ell =0}>0$, we need $(1-\frac{\beta_{\ell}}{2\pi r}) > 0$ or $r >\frac{\beta_{\ell}}{2\pi}$. This condition results along with the Eq.\eqref{eq:rho_l} for $r$,
\begin{align*}
\alpha^{\prime} \geqslant \frac{\beta_{\ell}}{2\pi} -\frac{1}{2\pi}\ln(\frac{2\pi \Lambda}{\beta_{\ell}}).
\end{align*} 

This stability condition for $r$ put a constraint on the mass term $\alpha^{\prime}$ as
\begin{equation}
 \alpha^{\prime} \geqslant \alpha^{\prime}_{0},
\end{equation} 
with 
\begin{align}\label{eq:alpha0}
\alpha^{\prime}_{0} =  \frac{\beta_{\ell}}{2\pi} -\frac{1}{2\pi}\ln(\frac{2\pi \Lambda}{\beta_{\ell}}).
\end{align} 
\begin{figure*}[t]
\begin{tabular}{@{}ccc@{}}
{\includegraphics[width=0.315\textwidth]{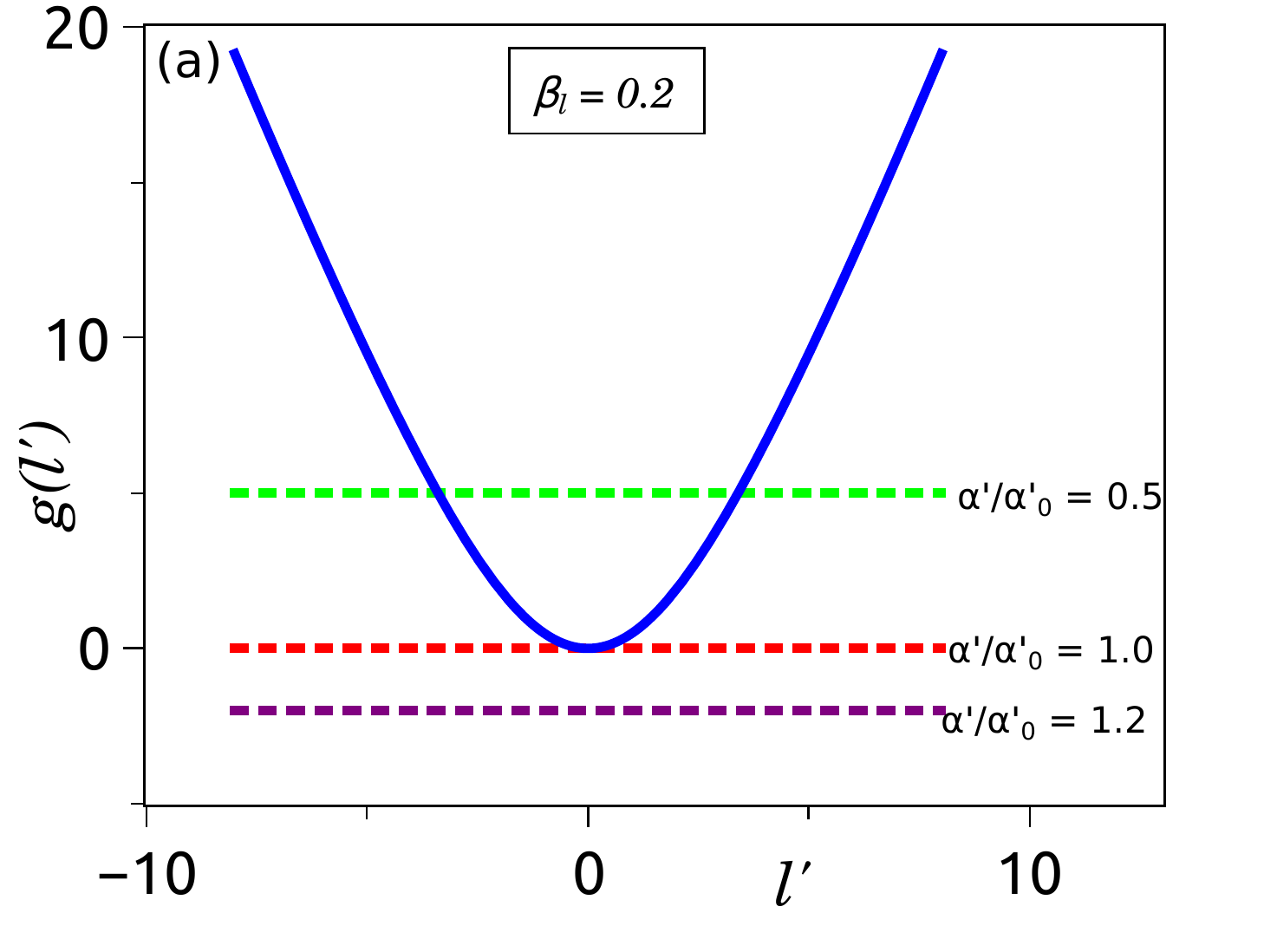}} &
{\includegraphics[width=.30\textwidth]{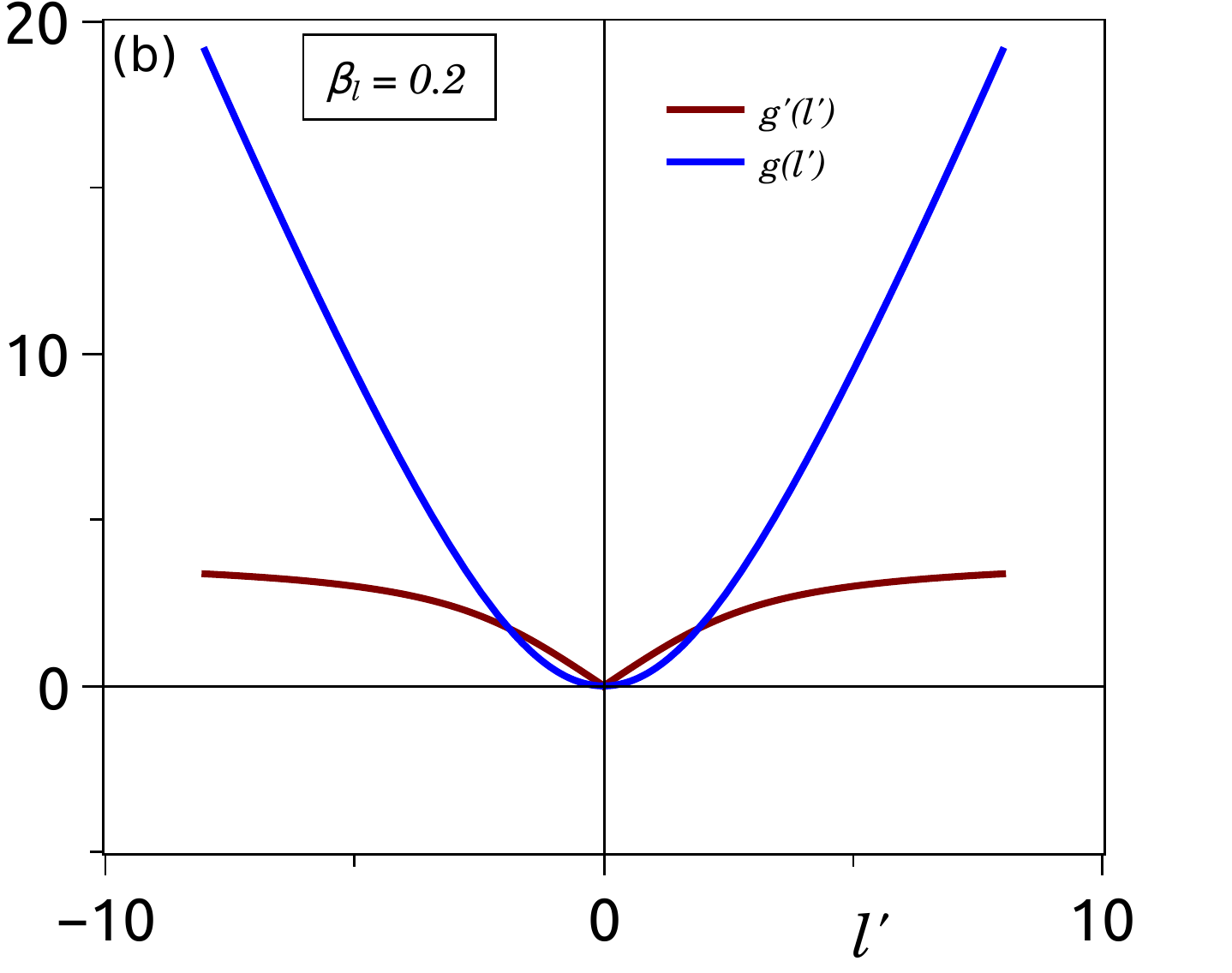}} &
{\includegraphics[width=.307\textwidth]{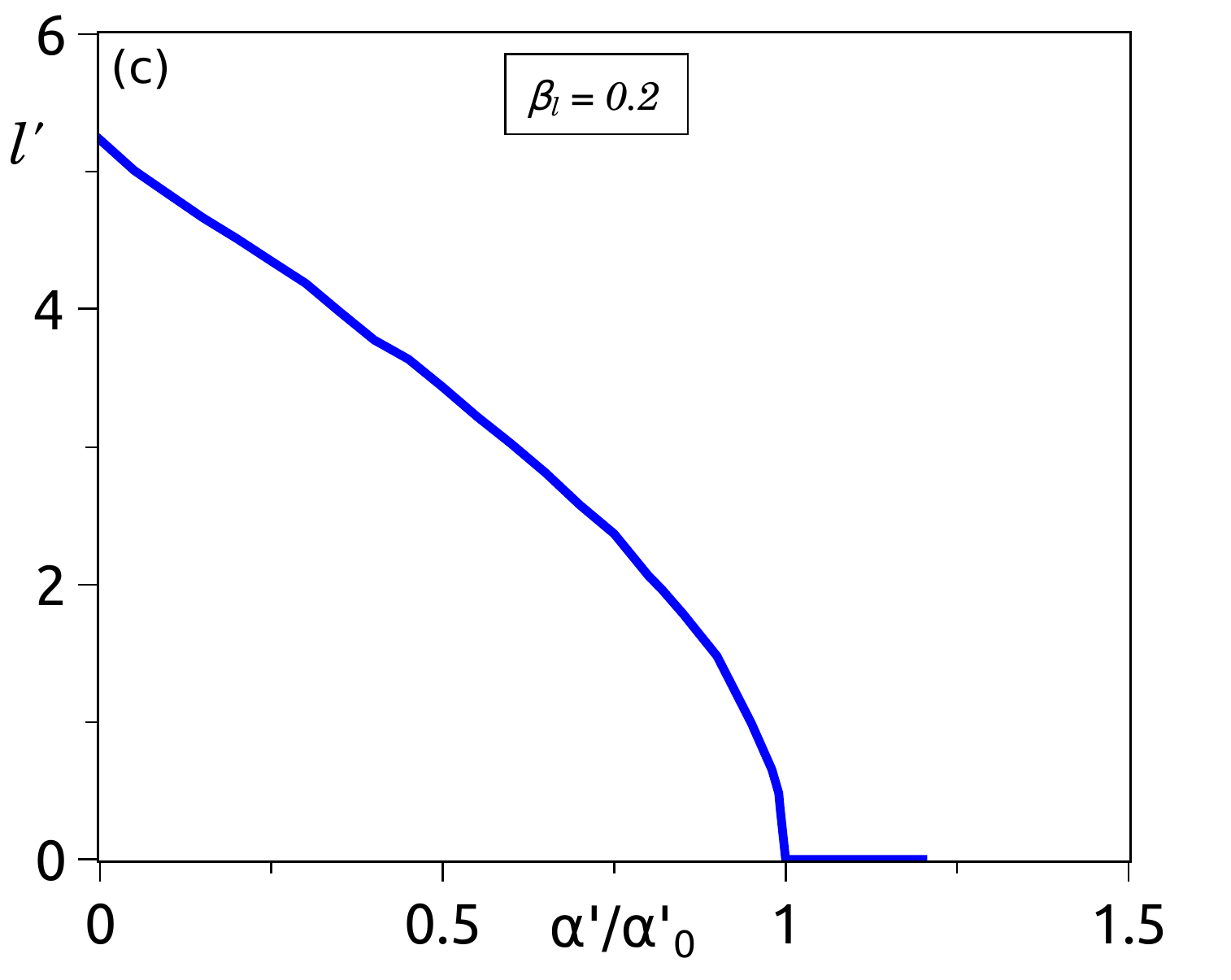}} \\
{\includegraphics[width=.315\textwidth]{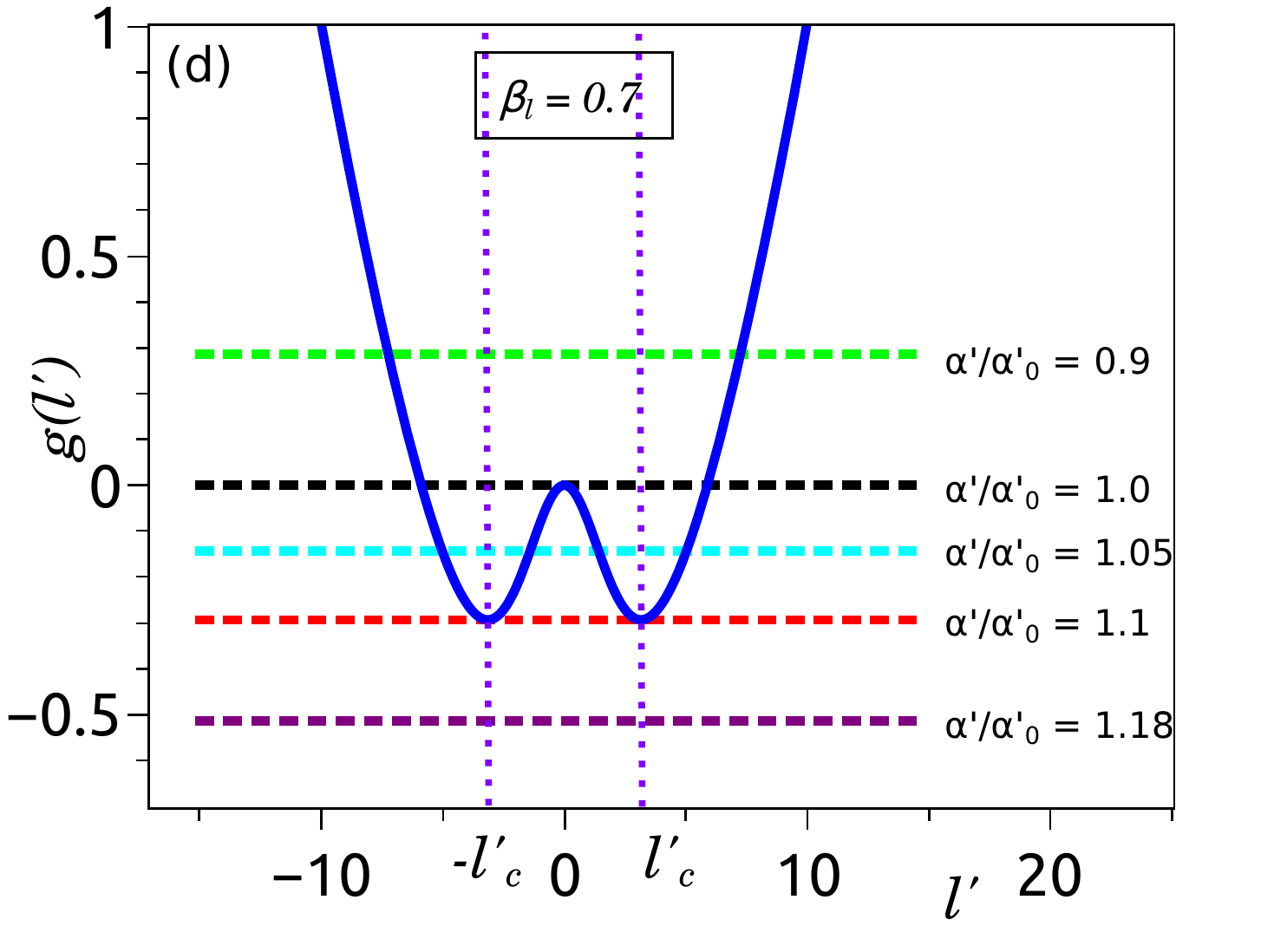}} &
{\includegraphics[width=.30\textwidth]{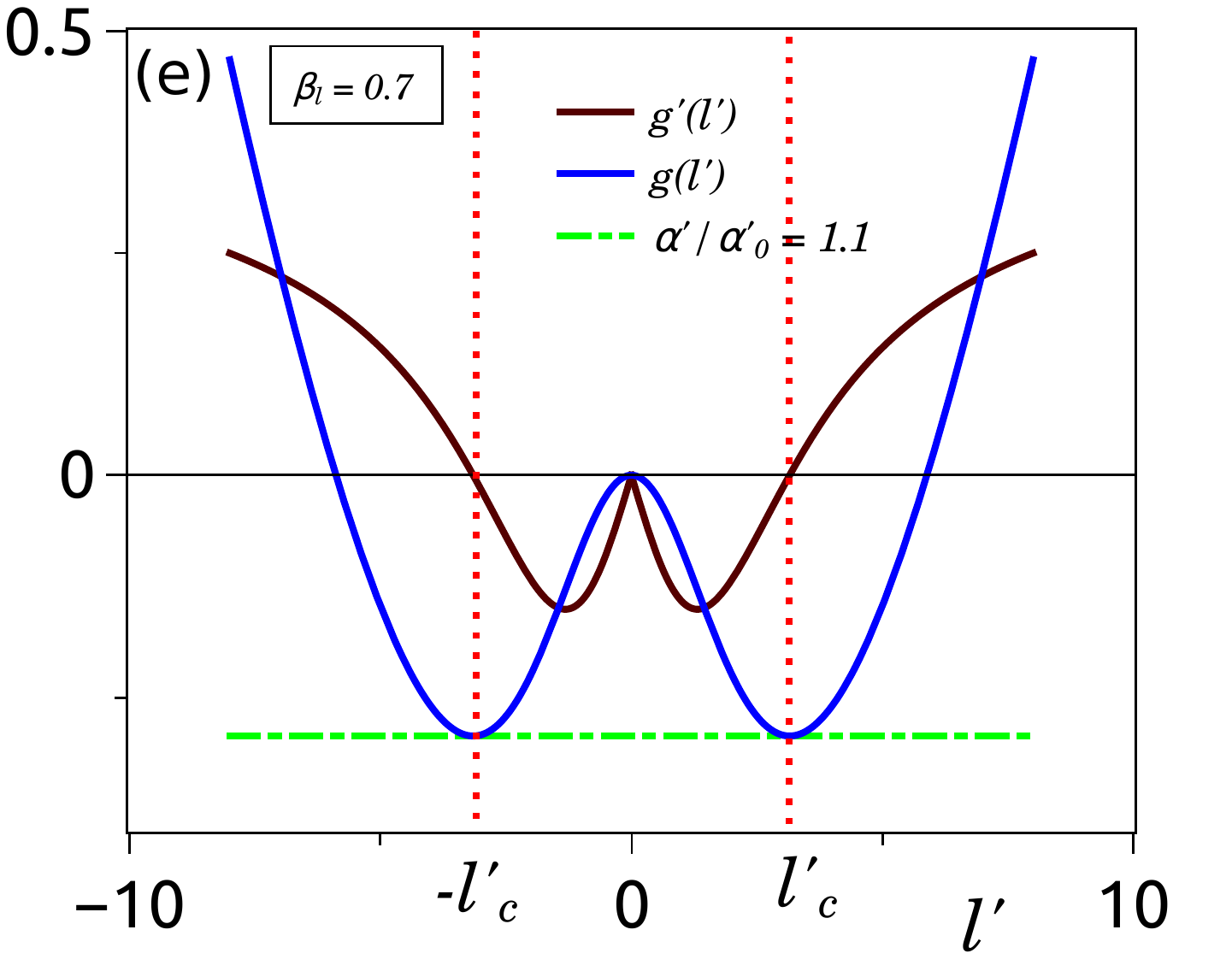}} &
{\includegraphics[width=.30\textwidth]{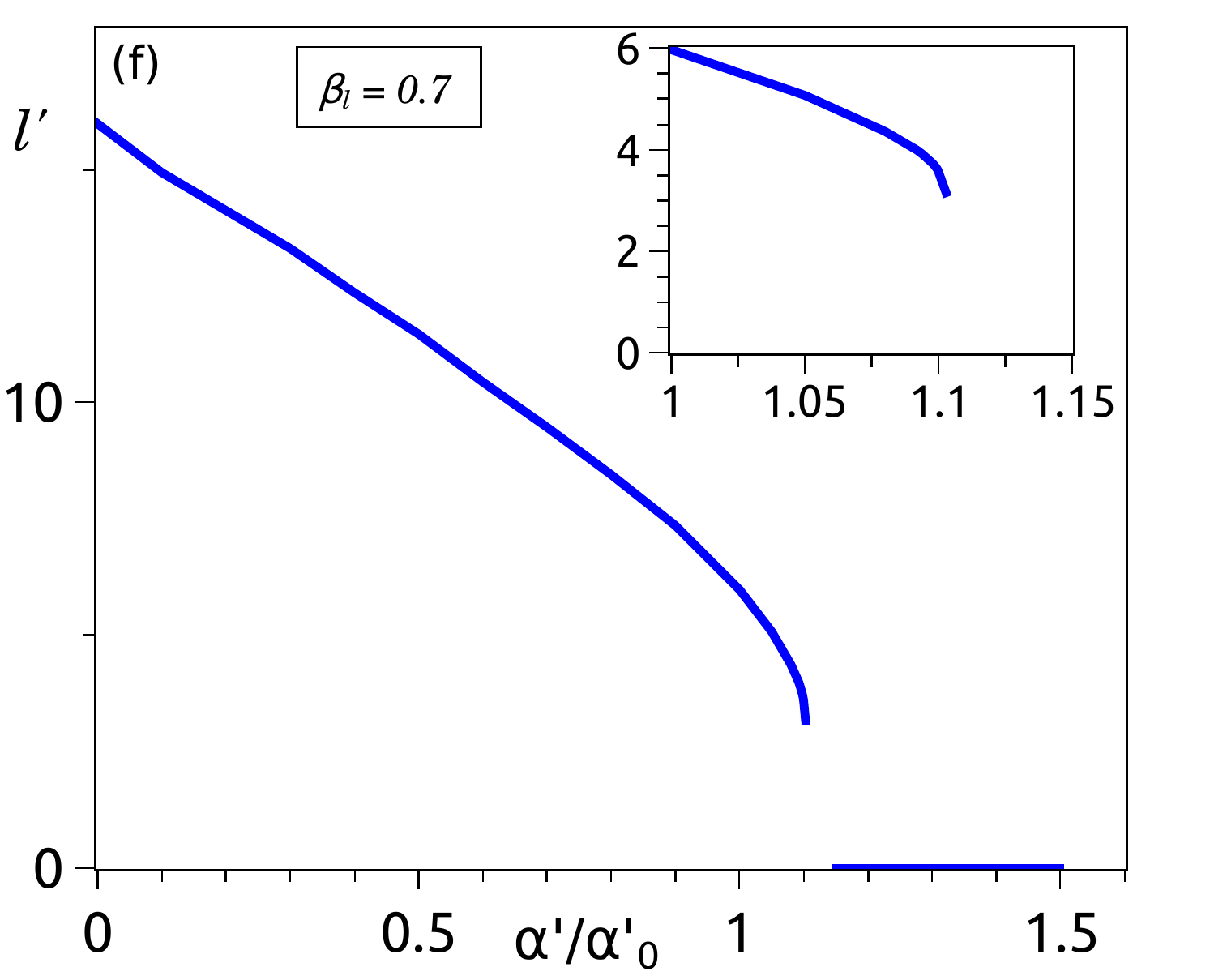}}
\end{tabular}
\caption{\label{Plotphase_trans}Graphical analysis illustrating the nature of phase transitions of the preemptive MELC order. In (a) and (d), the L.H.S. and R.H.S. of the Eq.\eqref{eq:func_l2} is shown as solid and dashed lines respectively. Intersections of the solid and the dashed lines give the solutions of $\ell^{\prime}$, where $\ell^{\prime}$ is the rescaled MELC order as given in the Eq.\eqref{eq:l_rescale}. The stability of the solutions is analyzed by looking at $g^{\prime}(\ell^{\prime})$ along with $g(\ell^{\prime})$ as shown in (b) and (e). The allowed stable solutions of $\ell^{\prime}$ are plotted in (c) and (f). The case of the second order phase transition is shown in (a-c) with a choice $\beta_{\ell} = 0.2$. $\ell^{\prime}$ continuously goes to zero at $\alpha^{\prime}/\alpha^{\prime}_{0} = 1$ [corresponding temperature is defined as T$_{0}$ in the Eq.\eqref{eq:T_alpha0}]. With the parameterization of $\alpha^{\prime}$ in Eq.\eqref{ew:Tfluc}, $\alpha^{\prime}=0$ corresponds to $T=T_{\text{PDW}}$, the mean-field transition temperature of the PDW field. Appearance of $\ell^{\prime}$ at T$_{0}$($>T_{\text{PDW}}$) shows that $\ell^{\prime}$ is a preemptive order. The case of the first order phase transition is shown in (d-f) with a representative $\beta_{\ell} = 0.7$. In contrast to the second order case, in this case, solutions of $\ell^{\prime}$ also exist for $\alpha^{\prime}/\alpha^{\prime}_{0} > 1$. Though there exist a parameter regime where there are four $\ell^{\prime}$ values, a closer investigation at $g^{\prime}(\ell^{\prime})$ in (e) show that the solutions in the range $-\ell^{\prime}_{c}<\ell^{\prime}<\ell^{\prime}_{c}$ are not stable, hence not allowed. This creates a discontinuity in the allowed values of $\ell^{\prime}$. For the parameters considered in this plot, the discontinuous jump in $\ell^{\prime}$ occurs at a value $\alpha^{\prime}/\alpha^{\prime}_{0}=1.1>1$ (corresponding temperature is T$_{0}^{\prime}$ and is also greater than $T_{\text{PDW}}$). In all the plots, we have taken $\pi \alpha^{\prime}_{0}=1$ for simplicity.}
\end{figure*}
\subsubsection{Case: $\ell \neq 0, \varrho \neq 0$}
Here we analyze the state where $\varrho \neq 0$ and $\ell\neq 0$, i.e. a state with preemptive MELC order. Eliminating $r$ from Eq.\eqref{eq:rho_l}, we arrive at the following equation for $\ell^{\prime}$,
\begin{align}\label{eq:l_state1}
\frac{\beta_{\ell}}{2\pi}\ell^{\prime}\coth\ell^{\prime} - \frac{1}{2\pi}\ln(\frac{2\pi\Lambda}{\beta_{\ell}}) +\frac{1}{2\pi}\ln(\frac{\ell^{\prime}}{\sinh\ell^{\prime}}) =\alpha^{\prime},
\end{align}
where
\begin{align}\label{eq:l_rescale}
\ell^{\prime}= \frac{2\pi \ell}{\beta_{\ell}}.
\end{align}
Substituting $\alpha^{\prime}_{0}$ from Eq.\eqref{eq:alpha0} in Eq.\eqref{eq:l_state1}, and rearranging we get,
\begin{align}\label{eq:func_l}
1 - \frac{\ell^{\prime}}{\tanh\ell^{\prime}} + \frac{1}{\beta_{\ell}}\ln\left(\frac{\sinh\ell^{\prime}}{\ell^{\prime}}\right) = \frac{2\pi}{\beta_{\ell}}\left(\alpha^{\prime}_{0} - \alpha^{\prime}\right).
\end{align}
The above equation gives the solution for $\ell^{\prime}$. We define the left hand side (L.H.S.) of the Eq.\eqref{eq:func_l} as $g(\ell^{\prime})$, i.e.
\begin{align}\label{eq:def_g}
 g(\ell^{\prime}) = 1 - \frac{\ell^{\prime}}{\tanh\ell^{\prime}} + \frac{1}{\beta_{\ell}}\ln\left(\frac{\sinh\ell^{\prime}}{\ell^{\prime}}\right).
\end{align} 
 We plot $g(\ell^{\prime})$ in Fig.  \ref{Plot_compare_betas} for different values of $\beta_{l}$.
  We notice that for $\ell^{\prime}>0$ and for $\beta_{\ell}<0.5$, the function $f(\ell^{\prime})$ is monotonically increasing, whereas for $1>\beta_{\ell}>0.5$, the function is not monotonically increasing. These are as well true for the $\ell^{\prime}<0$. We will show that the MELC order $\ell^{\prime}$ can appear through two types of phase transitions depending on the parameter regime $0.5>\beta_{\ell}>0$ and $1>\beta_{\ell}>0.5$.
\subsubsection*{Second order phase transition ($0.5>\beta_{\ell}>0$)}
First we discuss the case $0.5>\beta_{\ell}>0$. The right hand side (R.H.S.) of the Eq.\eqref{eq:func_l}, can be rewritten as $\frac{2\pi}{\beta_{\ell}}\alpha^{\prime}_{0}\left(1 - \alpha^{\prime}/\alpha^{\prime}_{0}\right)$, where $\alpha^{\prime}_{0}$ is also a function of $\beta_{\ell}$ and we consider $\alpha^{\prime}_{0}>0$. Hence Eq.\eqref{eq:func_l} becomes [using Eq.\eqref{eq:def_g}],
\begin{align}\label{eq:func_l2}
g(\ell^{\prime}) = \frac{2\pi \alpha^{\prime}_{0}}{\beta_{\ell}}\left(1 - \alpha^{\prime}/\alpha^{\prime}_{0}\right) .
\end{align}
The plot in the Fig. \ref{Plotphase_trans}(a) represents graphical representation of both sides of the Eq.\eqref{eq:func_l2} for $\beta_{\ell} = 0.2$  and three different values of $\alpha^{\prime}/\alpha^{\prime}_{0}$. We notice that for $ \alpha^{\prime}<\alpha^{\prime}_{0}$, the equation sustains non-zero values for $\ell^{\prime}$, while for $\alpha^{\prime} = \alpha^{\prime}_{0}$, $\ell^{\prime}$ becomes zero, and for $\alpha^{\prime}>\alpha^{\prime}_{0}$, Eq.\eqref{eq:func_l2} has no solution. So, the order $\ell^{\prime}$ appears first at $\alpha^{\prime} = \alpha^{\prime}_{0}$ and then increases as $\alpha^{\prime}/\alpha^{\prime}_{0}$ gets smaller. Whether the state with the values of $\ell^{\prime}$, obtained from solution of Eq.\eqref{eq:func_l2} is stable state or not can be seen by analyzing the second derivative of the effective free energy. The corresponding condition is $g^{\prime}(\ell^{\prime})|_{\ell^{\prime}=\ell_{0},-\ell_{0}} =\frac{\partial g(\ell^{\prime})}{\partial\ell^{\prime}}|_{\ell^{\prime}=\ell_{0},-\ell_{0}}>0$, where $\ell_{0}$ is a solution of the Eq.\eqref{eq:func_l2}. To analyze this condition, we determine  $g^{\prime}(\ell^{\prime})|_{\ell^{\prime}=\ell_{0}}$ and  $g^{\prime}(\ell^{\prime})|_{\ell^{\prime}=-\ell_{0}}$, which are given by the following equations,
{\small{
\begin{align}\label{eq:derivatives}
\nonumber
g^{\prime}(\ell^{\prime})|_{\ell^{\prime}=\ell_{0}} & = \frac{1}{\beta_{\ell}}\left[\frac{1}{\tanh\ell_{0}}-\frac{1}{\ell_{0}}\right] -\left[\frac{1}{\tanh\ell_{0}} -  \frac{\ell_{0}}{\sinh^{2}\ell_{0}}\right] \\
g^{\prime}(\ell^{\prime})|_{\ell^{\prime}=-\ell_{0}} & = \frac{1}{\beta_{\ell}}\left[\frac{-1}{\tanh\ell_{0}} + \frac{1}{\ell_{0}}\right] -\left[\frac{-1}{\tanh\ell_{0}} + \frac{\ell_{0}}{\sinh^{2}\ell_{0}}\right].
\end{align}}}
$g^{\prime}(\ell^{\prime})$ and $g(\ell^{\prime})$ for $\beta_{\ell} = 0.2$ are plotted in the Fig. \ref{Plotphase_trans}(b). We observe that in this case, $g^{\prime}(\ell^{\prime})>0$ for all $\ell^{\prime}$. Hence all the solutions of $\ell^{\prime}$ from Eq.\eqref{eq:func_l2} are allowed and correspond to the minima of the effective free energy.

The allowed values of $\ell^{\prime}$ is plotted as a function of $\alpha^{\prime}/\alpha^{\prime}_{0}$ for $\beta_{\ell} = 0.2$ in Fig. \ref{Plotphase_trans}(c). We notice that value of $\ell^{\prime}$ continuously decreases to zero as $\alpha^{\prime}/\alpha^{\prime}_{0}$ approaches 1. This indicates a second order phase transition in the MELC order $\ell^{\prime}$.

\subsubsection*{First order phase transition ($1>\beta_{\ell}>0.5$)}
Here we discuss the case where $1>\beta_{\ell} > 0.5$. Again, we vary $\alpha^{\prime}/\alpha^{\prime}_{0}$ to find the solution of the Eq.\eqref{eq:func_l2}. We plot the L.H.S. and R.H.S. of Eq.\eqref{eq:func_l2} for $\beta_{\ell} = 0.7$  in Fig. \ref{Plotphase_trans}(d). We observe that in this case, as $\alpha^{\prime}/\alpha^{\prime}_{0}$ is increased from zero to one, the value of $\ell^{\prime}$ is non-zero and it decreases as in the case of $\beta_{\ell} <0.5$. But Eq.\eqref{eq:func_l2} also has solution for $\alpha^{\prime}/\alpha^{\prime}_{0}>1.0$, which is in striking contrast to the $\beta_{\ell} <0.5$ case. We also notice that for $\alpha^{\prime}/\alpha^{\prime}_{0}>1.0$, the Eq.\eqref{eq:func_l2} has two solutions. To analyze whether both the solutions are stable, we plot $g^{\prime}(\ell^{\prime})$ and $g(\ell^{\prime})$ in the Fig. \ref{Plotphase_trans}(e) for the case $\beta_{\ell} = 0.7$. We observe that for $\ell^{\prime}> \ell^{\prime}_{c}$, as indicated by the red dotted line, $g^{\prime}(\ell^{\prime})>0$, whereas for $\ell^{\prime}< \ell^{\prime}_{c}$, $g^{\prime}(\ell^{\prime})<0$. This implies that all the values of $\ell^{\prime}>\ell^{\prime}_{c}$ are stable and therefore correspond to minima of the free energy. Hence, for the case $1>\beta_{\ell}>0.5$, the allowed values of $\ell^{\prime}$ remains finite from $\alpha^{\prime}/\alpha^{\prime}_{0}=0$ till a certain value of $\alpha^{\prime}/\alpha^{\prime}_{0}(>1)$, and then suddenly jumps to zero as beyond that particular $\alpha^{\prime}/\alpha^{\prime}_{0}$, there exists no solution to the Eq.\eqref{eq:func_l2}. 

In Fig. \ref{Plotphase_trans}(f), we plot the allowed values of $\ell^{\prime}$ as a function of $\alpha^{\prime}/\alpha^{\prime}_{0}$. We notice the discontinuous jump in $\ell^{\prime}$ at a certain value of $\alpha^{\prime}/\alpha^{\prime}_{0}(>1)$. This discontinuous change in the $\ell^{\prime}$ suggests a first-order phase transition. 

\subsubsection*{Temperature dependence of $\ell^{\prime}$: Preemptive MELC order}
\begin{figure}[t]
\includegraphics[width=0.8\linewidth]{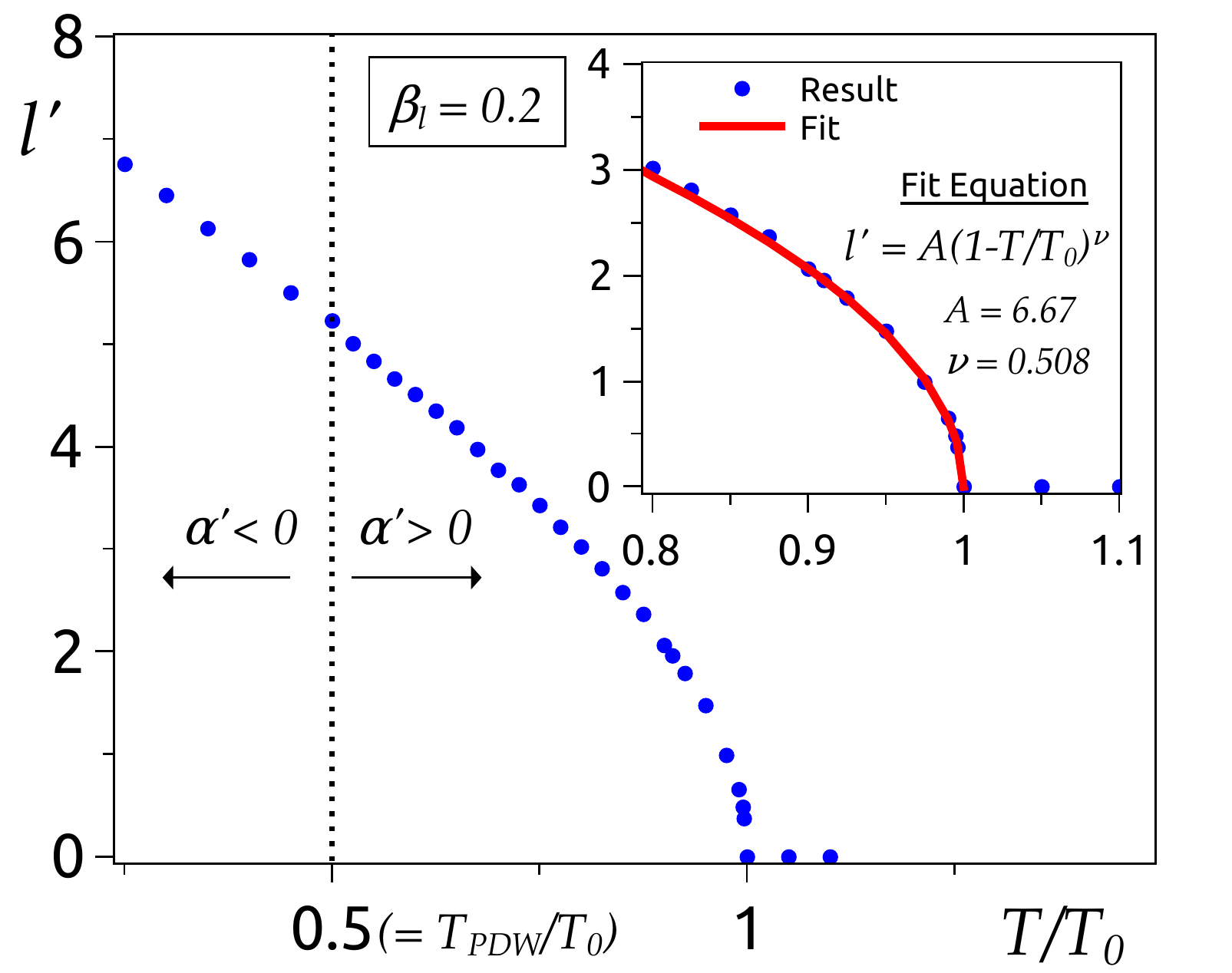}
\caption{\label{Plot:LTphase}A phase diagram of preemptive MELC order ($\ell^{\prime}$) [defined in Eq.\eqref{eq:l_rescale}] with scaled temperature $T/T_{0}$ for the case of second order phase transition. Values of $\ell^{\prime}$ are obtained from Fig.\ref{Plotphase_trans}(c) and the temperature dependence is calculated using the parameterization of $\alpha^{\prime}$ in Eq.\eqref{ew:Tfluc}. To plot this phase diagram we have taken $T_{0} = 1$ and $T_{\text{PDW}}/T_{0} = 0.5$ and $\beta_{\ell} = 0.2$. The plot shows $\ell^{\prime}$ continuously changes across T$_{\text{PDW}}$ where $\alpha^{\prime}$ changes it's sign. This indicates that the preemptive MELC order first emerges at temperature $T=T_{0}$ and can survive down to low temperatures. The inset figure shows a fit of the values of $\ell^{\prime}$ in the main panel close to $T=T_{0}$ with a function $A(1 - T/T_{0})^\nu$. The fitting parameter $\nu$ is found to be $\approx0.5$, close to mean-field critical exponent in Ising like transition. Similar temperature dependence can be found directly from a low order expansion of $\ell^{\prime}$ in Eq.\eqref{eq:func_l} which yields Eq.\eqref{eq:l_lowT}. }
\end{figure}
To study the temperature dependence of $\ell^{\prime}$, we parameterize $\alpha^{\prime}$ as,
\begin{align}\label{ew:Tfluc}
\alpha^{\prime} = M^{\prime}(T- T_{\text{PDW}}),
\end{align} 
where $M^{\prime}$ is a positive constant. The parameterization of $\alpha^{\prime}$ is chosen in such a way that there is a mean-field phase transition at temperature $T = T_{\text{PDW}}$ to a composite PDW order  i.e. when $\alpha^{\prime} = 0$. 
 Therefore $\alpha^{\prime}$ corresponding to some arbitrary temperature $T_{0}$ can be written as
\begin{align}\label{eq:T_alpha0}
\alpha^{\prime}\vert_{T = T_{0}} \equiv \alpha^{\prime}_{0} = M^{\prime}(T_{0} - T_{\text{PDW}}) .
\end{align}
This leads to 
\begin{align}\label{eq:T_0}
T_{0} &= T_{\text{PDW}} + \frac{\alpha^{\prime}_{0}}{M^{\prime}}.
\end{align}

We note that $\alpha^{\prime}> \alpha^{\prime}_{0}$ corresponds to temperature $T > T_{0}$ and $\alpha^{\prime}< \alpha^{\prime}_{0}$ to temperature $T< T_{0}$. Since $\alpha^{\prime}_{0}>0$, we observe that T$_{0}$ is always larger than T$_{\text{PDW}}$, which is the condensation temperature for the PDW fields. The MELC order appears through a continuous second-order phase transition for $0.5>\beta_{\ell}>0$, at temperature $T_{0}$, which is higher than $T_{\text{PDW}}$. This indicates that the order $\ell^{\prime}$ preempts the PDW order, hence the MELC order appears as a preemptive order in our analysis.
As noted in Eq.\eqref{ew:Tfluc}, $\alpha^{\prime}/\alpha^{\prime}_{0}$ is directly proportional to the temperature. Consequently, for the first-order phase transition for $1.0>\beta_{\ell}>0.5$, the MELC order appears at a temperature $T^{\prime}_{0}$. $T^{\prime}_{0}$ is even higher than the second order transition temperature $T_{0}$ and certainly greater than $T_{\text{PDW}}$ [see Eq.\eqref{eq:T_0}]. Thus, the MELC order preempts $T_{\text{PDW}}$ even in this case of first order phase transition.
 Within our theoretical framework, the PDW order is a composite order of SC and BDW. Hence it becomes long-range only when both the primary orders SC and BDW become long-range. This suggests that the MELC order also preempts the primary SC and BDW orders. It is important to note that $\mathcal{T}-\mathcal{P}$ symmetry breaking has been experimentally observed at pseudo-gap temperature T$^{*}$ in the under-doped cuprates. The temperature T$_{0}$ can as well be T$^{*}$. However to establish whether the temperature T$_{0}$ is equal to T$^{*}$ is beyond the scope of this work.

For the case of second-order transition, we extract the temperature dependence of $\ell^{\prime}$ by obtaining the solutions of $\ell^{\prime}$ in Fig. \ref{Plotphase_trans}(c) and using the temperature parameterization of $\alpha^{\prime}$ in Eq.\eqref{ew:Tfluc}. In Fig. \ref{Plot:LTphase}, we plot the temperature dependence of $\ell^{\prime}$ for $\beta_{\ell}=0.2$. Close to the transition, an analytical temperature dependence of $\ell^{\prime}$ can be obtained. If we expand the L.H.S. of the Eq.\eqref{eq:func_l2} for small values of $\ell^{\prime}$, we find that
\begin{align}
\frac{\ell^{\prime 2}}{ 6\beta_{\ell}} = \frac{2\pi \alpha^{\prime}_{0}}{\beta_{\ell}}\left(1 - \alpha^{\prime}/\alpha^{\prime}_{0}\right) .
\end{align}
This gives,
\begin{align}
\ell^{\prime 2} = 12 \pi (\alpha_{0}^{\prime}-\alpha^{\prime}) 
\end{align}
or using Eq.\eqref{ew:Tfluc} and Eq.\eqref{eq:T_alpha0}, 
\begin{align}\label{eq:l_lowT}
\ell^{\prime } \propto (T_{0}-T)^{1/2}.
\end{align}

Since we have only obtained the solutions for $\ell^{\prime}$ within a HS saddle-point analysis, the fluctuations in $\ell^{\prime}$ are not captured in our formalism. Thus, the critical exponent obtained here behaves similar to a mean-field Ising like transition. Fluctuations of $\ell^{\prime}$ can be considered within an alternate method like the renormalization group approach. Such an analysis to obtain a non-mean-field like exponent is not a part of this study.

In the inset of Fig. \ref{Plot:LTphase}, we show the temperature dependence of $\ell^{\prime}$ close to the transition. Note that the points in this inset are obtained from the main panel and thus satisfy Eq.\eqref{eq:func_l}. As an independent check, we fit $\ell^{\prime}$ for temperatures close to T$_{0}$ with a function $A(1 - T/T_{0})^\nu$, where $A$ and $\nu$ are fitting parameters. The value of $\nu$ obtained from the fit is consistent with the analytic expression of the temperature dependence in Eq.\eqref{eq:l_lowT}.

\section{Discussions}\label{Sec_Disc}

\subsection{Extension to tetragonal lattice systems}\label{sec_om}
In order to construct the GL theory [see Sec.\ref{sec:GL_nofluc} and Sec\ref{sec:GL_fluc}], we restrict ourselves to lattice systems having orthorhombic symmetry to simplify the analytic calculations. However, the GL framework presented in this paper can be extended to larger order parameter manifold, relevant to lattice structure with higher symmetries.

To demonstrate, we construct the order parameter manifold and MELC order for tetragonal lattice systems with $C_{4}$ symmetry. For orthorhombic system, the number of relevant hot-spots are two [numbered as 1 and 5 in Fig. \ref{Fig_FS}] as described in Sec \ref{symmetry_op}. The order parameter manifold is given in the Eq.\eqref{eq:sigmay_2}. We need seven unknown GL parameters to write down the corresponding free energy in Eq.\eqref{eq_GLFE} for this reduced order parameter manifold. For a tetragonal crystal system, due to presence of $C_{4}$ symmetry, the relevant hot-spots are 1, 3, 5 and 7. The order parameter manifold in this case becomes
\begin{align}\label{eq:OM_4}
[\chi_{Q_{x}}^{1},\chi_{Q_{y}}^{3},\chi_{-Q_{x}}^{5},\chi_{-Q_{y}}^{7},\Delta].
\end{align}
The GL free energy in this case can be written following the similar procedure as in section \ref{sec:GL_nofluc} as,
\begin{align}\label{eq:free_tetra}
\begin{split}
\mathcal{F} &= \alpha_{d}|\Delta|^{2} + \alpha(|\chi_{Q_{x}}^{1}|^{2} + |\chi_{Q_{y}}^{3}|^{2} + |\chi_{-Q_{x}}^{5}|^{2} + |\chi_{-Q_{y}}^{7}|^{2}) \\
& + \beta_{1}(|\chi_{Q_{x}}^{1}|^{4} + |\chi_{Q_{y}}^{3}|^{4} + |\chi_{-Q_{x}}^{5}|^{4} + |\chi_{-Q_{y}}^{7}|^{4})\\
& + \beta_{d}|\Delta|^{4} + \beta_{2}(|\chi_{Q_{x}}^{1}|^{2}|\Delta|^{2} + |\chi_{-Q_{x}}^{5}|^{2}|\Delta|^{2} + |\chi_{Q_{y}}^{3}|^{2}|\Delta|^{2} \\
& + |\chi_{-Q_{y}}^{7}|^{2}|\Delta|^{2}) + \beta_{3}(|\chi_{Q_{x}}^{1}|^{2}|\chi_{-Q_{x}}^{5}|^{2} + |\chi_{Q_{y}}^{3}|^{2}|\chi_{-Q_{y}}^{7}|^{2})\\
& + \beta_{5}(|\chi_{Q_{x}}^{1}|^{2}|\chi_{Q_{y}}^{3}|^{2} + |\chi_{-Q_{x}}^{5}|^{2}|\chi_{-Q_{y}}^{7}|^{2} + |\chi_{Q_{y}}^{3}|^{2}|\chi_{-Q_{x}}^{5}|^{2} \\
& + |\chi_{Q_{x}}^{1}|^{2}|\chi_{-Q_{y}}^{7}|^{2}).
\end{split}
\end{align}
Here also, we notice that the free energy Eq.\eqref{eq:free_tetra} remains invariant under all point group operations of tetragonal system, parity, time-reversal, translation and gauge symmetry transformations. While writing the free energy in Eq.\eqref{eq:free_tetra}, we have not shown the terms which give rise to purely imaginary BDW ground states for simplicity. Enhanced number of BDW order parameters in the manifold [Eq.\eqref{eq:OM_4}] for tetragonal case gives increased number of possible mean-field solutions compared to the orthorhombic case. Furthermore, the number of unknown GL parameters is also increased in Eq.\eqref{eq:free_tetra}. So, the analysis of even the various mean-field solutions of Eq.\eqref{eq:free_tetra} becomes analytically cumbersome.

\subsection{Possible relevance to experiments}\label{Sec_Exp}

In this section we present possible outcomes of our analysis in connection to experimental observations in under-doped cuprates. Though our analysis of $\mathcal{T-P}$ symmetry breaking is based on phenomenological motivations drawn from experiments in under-doped cuprates, here we focus only on qualitative relevance to experiments due to the simplistic nature of the crystal structure considered in this paper.

We demonstrated in section \ref{sec:GL_fluc}, that the preemptive MELC order can appear through two types of phase transitions. First, it can appear through a continuous second-order phase transition at a temperature higher than T$_{\text{PDW}}$. Importantly, the phase transition temperatures can as well be equal to T$^{*}$. The signatures of appearance of IUC magnetism breaking $\mathcal{T-P}$ symmetry at T$^{*}$ through a second-order transition have been reported in some spin-polarized neutron diffraction experiments \cite{Tang_18,zhang18,Pal18,Itoh17}. Second, the MELC order can also appear through a discontinuous first-order phase transition at a temperature which is again higher than T$_{\text{PDW}}$ and even higher than the second-order phase transition temperature. To the best of our knowledge, there has been no experiment indicating a first order phase transition to a $\mathcal{T}-\mathcal{P}$ broken state.

Spin-polarized neutron scattering experiments measure the magnetic neutron intensity to describe the $\vec{Q}=0$ magnetism in the PG phase of cuprates. The observations allowed to deduce a critical exponent corresponding to the temperature dependence of the magnetic scattering intensity, although varying in a wide range of values from 0.25 to 0.5 \cite{Mook08,Li11,Tang_18}. Within our framework, in the case of the second order phase transition in sec. \ref{Saddle_sub1}, the temperature dependence of the preemptive MELC order $\ell^{\prime}$ close to the phase transition is obtained to be $\ell^{\prime} \sim (T_0-T)^{1/2}$ [Eq.\eqref{eq:l_lowT}]. This yields a critical exponent of $\ell^{\prime}$ to be 0.5. But it must be noted that a quantitative comparison of the exponents would require more accuracy in experimental results and also the consideration of fluctuations in $\ell^{\prime}$ using other theoretical tools like the renormalization group treatment.

For T$<$T$_{\text{PDW}}$, the GL parameter $\alpha^{\prime}<0$ as seen from Eq.\eqref{ew:Tfluc}. Remarkably even for $\alpha^{\prime}<0$, Eq.\eqref{eq:func_l2} has allowed solutions for both the cases $0.5>\beta_{\ell}>0$ and $1>\beta_{\ell}>0.5$. Therefore, the MELC order $\ell^{\prime}$ has non-zero value for $\alpha^{\prime} < 0$ and continuously changes when $\alpha^{\prime}$ becomes greater than zero. This shows that the preemptive MELC order persists below the temperature T$_{\text{PDW}}$. The Fig. \ref{Plot:LTphase} shows the existence of MELC order at low temperatures for the second order phase transition. This is also true for the first order phase transition. Although no signature of MELC order has been reported at low temperatures in the superconducting state for technical issues so far. This will motivate further experiments investigating $\mathcal{T-P}$ symmetry breaking at low temperatures in the superconducting state. 

It is also interesting to discuss the effects of the impurities on the MELC state. Strong substitutional impurities like Zn destroys the superconducting order parameter locally. Local MELC order parameter is given as $\ell \propto |\Phi_{i}|^2 \propto |\Delta_{ij}|^2|\chi_{ij}|^2$, where $\Delta_{ij}$ and $\chi_{ij}$ are superconducting and BDW order parameters respectively. The MELC order parameter is thus suppressed close to the impurities. As a result, the MELC order parameter is reduced with increase in Zn concentration. This might give an explanation to the reduction in the intensity of the IUC signal in polarized neutron diffraction measurement with Zn doping \cite{Baledent11}.

\section{Conclusion}\label{conclusion}
Considering SC and BDW as primary orders, in this paper, we explored possibility of $\mathcal{T-P}$ symmetry breaking in the PG state of under-doped cuprates. We found that the thermal fluctuations of SC and BDW fields and consequently the fluctuations in the composite PDW fields result into a $\mathcal{T-P}$ symmetry breaking preemptive MELC ground state.

As a first step, within the GL mean-field theory of competing primary BDW and SC orders, we explored the existence of the MELC order in various parameter regimes and presented the conditions for the stability of the mean-field MELC state. We showed that the mean-field MELC order can emerge only in a phase where SC and BDW coexist. This mean-field MELC order is restricted to a situation where the BDW ground state itself breaks $\mathcal{T-P}$. However, there is no experimental evidence of such a BDW ground state.

We then considered the fluctuations of the SC and BDW fields in the free energy. We notice that fluctuations in SC and BDW fields lead to fluctuations in composite PDW fields. Treating the thermal fluctuations of the PDW fields in a Hubbard Stratonovich approach, we then showed that a nontrivial MELC order preempts at a temperature above T$_{\text{PDW}}$, the mean-field PDW transition temperature. The analysis showed that the $\mathcal{T-P}$ symmetry is spontaneously broken in the preemptive MELC state even though the PDW ground states preserve the symmetry. This also suggests that the BDW ground state does not need to break $\mathcal{T-P}$ symmetry. We described that, depending on parameter regime, the preemptive MELC order can emerge through both second and first order phase transition.

\begin{acknowledgments}

We thank Y. Sidis for valuable discussions. This work has received financial support from the ERC, under grant agreement AdG-694651-CHAMPAGNE.

\end{acknowledgments}

\appendix
\section{Details of mean-field GL theory}\label{sec:AppA}
In this Appendix, we analytically calculate the mean-field solutions of the free energy density Eq.\eqref{eq_GLFE} for competing BDW and SC. The possible ground states for the primary order parameters ($\chi_{Q}, \chi_{-Q}, \Delta$) manifold are shown in the table \ref{tstaes}. For each ground state, we calculate the mean-field free energy, which enables us to find the stability criteria of the ground state sustaining a non-zero MELC order.
\subsection{ Ground state (a, 0, b)}
The free energy density functional for this state can be obtained from Eq.\eqref{eq_GLFE} and is given by,
\begin{equation}\label{eq:fr_c1}
F = \alpha_{d}|\Delta|^{2} + \alpha |\chi_{Q}|^{2} + \frac{\beta_{1}}{2}|\chi_{Q}|^{4} + \frac{\beta_{d}}{2}|\Delta|^{4} + \beta_{2}|\chi_{Q}|^{2}|\Delta|^{2}.
\end{equation}
Minimization of the free energy with respect to $\chi_{Q}$ and $\Delta$ leads to the following two mean-field equations:
\begin{equation}
 \alpha + \beta_{1} |\chi_{Q}|^{2} + \beta_{2}|\Delta|^{2} = 0
\end{equation}
 and,
\begin{equation}
\alpha_{d} + \beta_{d} |\Delta|^{2} + \beta_{2}|\chi_{Q}|^{2} = 0. 
\end{equation}
Above two equations give the mean-field values of $\chi_{Q}$ and $\Delta$ to be,
\begin{equation}
\begin{split}
\chi_{Q}^{2} = \frac{\beta_{2}\alpha_{d}-\alpha \beta_{d}}{\beta_{1}\beta_{d} - \beta_{2}^{2}}, \\
\Delta^{2} = \frac{\alpha \beta_{2}- \alpha_{d}\beta_{1}}{\beta_{1}\beta_{d} - \beta_{2}^{2}}.
\end{split}
\end{equation}

Substituting the above solution in the mean-field free energy, we get,
\begin{equation} \label{ew:fr_fc1}
F (a,0,b) = \frac{\left[2\alpha\alpha_{d}\beta_{2}- \alpha_{d}^{2}\beta_{1}- \alpha^{2}\beta_{d}\right]}{2\left[\beta_{1}\beta_{d} - \beta_{2}^{2}\right]}.
\end{equation}
\subsection{Ground state (a, a, b)}
The free energy density functional in this case can be written from Eq.\eqref{eq_GLFE} as,

\begin{equation}\label{eq:fr_c2}
\begin{split}
F = \alpha_{d}|\Delta|^{2} + 2\alpha |\chi_{Q}|^{2} + \beta_{1}|\chi_{Q}|^{4} + \frac{\beta_{d}}{2}|\Delta|^{4}\\ + 2\beta_{2}|\chi_{Q}|^{2}|\Delta|^{2} +
\beta_{3}|\chi_{Q}|^{4} + 2\beta_{4}|\chi_{Q}|^{2}|\Delta|^{2}.
\end{split}
\end{equation}
Minimizing the free energy Eq.\eqref{eq:fr_c2} w.r.t. $\chi_{Q}$ and $\Delta$ gives,
\begin{equation}
\alpha + \chi_{Q}^{2} \left(\beta_{1} + \beta_{3}\right) + \Delta^{2}\left(\beta_{2} + \beta_{4}\right) = 0
\end{equation}
and
\begin{equation}
\alpha_{d} + 2(\beta_{2} + \beta_{4})\chi_{Q}^{2} + \beta_{d}\Delta^{2} = 0.
\end{equation}

Above two equations give the mean-field solution as,
\begin{equation}
\begin{split}
\chi_{Q}^{2} =\frac{\alpha_{d}(\beta_{2}+\beta_{4})- \alpha \beta_{d}}{\beta_{d}(\beta_{1} + \beta_{3}) - 2(\beta_{2}+\beta_{4})^{2}},\\
\Delta^{2} = \frac{2\alpha(\beta_{2}+\beta_{4})- \alpha_{d} (\beta_{1} + \beta_{3})}{\beta_{d}(\beta_{1} + \beta_{3}) - 2(\beta_{2}+\beta_{4})^{2}}.
\end{split}
\end{equation}
The mean-field free energy corresponding to this solution is,
\begin{equation}
F(a,a,b) = \frac{4\alpha\alpha_{d}(\beta_{2} + \beta_{4}) - \alpha_{d}^{2}(\beta_{1}+\beta_{3})-2\alpha^{2}\beta_{d}}{2\beta_{d}(\beta_{1}+\beta_{3})-4(\beta_{2}+\beta_{4})^{2}}.
\end{equation}
\subsection{Ground state (a, 0, 0)}
The free energy density Eq.\eqref{eq_GLFE} in this state can be written as,
\begin{equation} \label{eq:fr_c4}
F = \alpha|\chi_{Q}|^{2} + \frac{\beta_{1}}{2}|\chi_{Q}|^{4}.
\end{equation}
Minimizing the above free energy w.r.t. $\chi_{Q}$ gives,
\begin{equation}
\chi_{Q}^{2} = \frac{-\alpha}{\beta_{1}}.
\end{equation}
The mean-field free energy corresponding to this state is given by,
\begin{equation}
F (a,0,0) = \frac{-\alpha^{2}}{2\beta_{1}}.
\end{equation}
\subsection{Ground state (0, 0, b)}
The free energy Eq.\eqref{eq_GLFE} in this state is given by,
\begin{equation} \label{eq:fr_c6}
F = \alpha_{d}|\Delta|^{2} + \\
\frac{\beta_{d}}{2}|\Delta|^{4}.
\end{equation}
Minimizing the above free energy w.r.t. $\Delta$ gives,
\begin{equation}
\Delta^{2} = \frac{-\alpha_{d}}{\beta_{d}}.
\end{equation}

The corresponding mean-field free energy for the ground state is,
\begin{equation}
F(0,0,b) = \frac{-\alpha_{d}^{2}}{2\beta_{d}}.
\end{equation}
\subsection{Ground state (a, a, 0)}
Finally we consider the case, where only BDW is non-zero and no SC is present.

The free energy in Eq.\eqref{eq_GLFE} becomes,
\begin{align}\label{eq:fr_c7}
F = 2\alpha|\chi_{Q}|^{2} + \beta_{1}|\chi_{Q}|^{4} + \beta_{3}|\chi_{Q}|^{4}.
\end{align}
Minimizing the free energy w.r.t. $\chi_{Q}$ gives
\begin{equation}
\chi_{Q}^{2} = \frac{-\alpha}{(\beta_{1} + \beta_{3})}.
\end{equation}

The mean-field free energy for this ground state is,
\begin{equation}
F(a,a,0) = \frac{-\alpha^{2}}{(\beta_{1} + \beta_{3})}.
\end{equation}

\pagebreak

\bibliographystyle{apsrev4-1}
\bibliography{Cuprates}
\end{document}